# XGBoost-Powered Digital Twins Leverage Routine Blood Tests for Early Detection of Cancer and Cardiovascular Disease

**Lo Kai Shun John**[1,4], **Riya Nagar**[1,2,3,4], **Abicumaran Uthamacumaran**[3], **Hector Zenil**[1,2,3,4*]

[1]Algorithmic Dynamics Lab, King's Institute for Artificial Intelligence, King's College London, UK
[2] Cancer Research Interest Group, The Francis Crick Institute, London, UK
[3] Oxford Immune Algorithmics, Oxford University Innovation & London Institute for Healthcare Engineering, UK
[4] Departments of Biomedical Computing and Digital Twins, School of Biomedical Engineering and Medical Sciences, King's Faculty of Life Sciences and Medicine, King's Health Partners AHSC (NHS–KCL), King's College London, UK

## ABSTRACT

Early detection of cancer and cardiovascular diseases is fundamental to improving patient outcomes and reducing healthcare expenditure. Current cancer screening programmes are targeted towards specific cancers and are often inaccessible to large parts of the population, particularly in remote regions. This project aimed to develop *digital blood twins*: machine learning models that leverage routinely collected blood test data, demographics, comorbidities, and prescribed medications, for scalable and cost-effective disease screening. Digital blood twins were constructed using the UK Biobank dataset (n = 373,269). Using age, sex, comorbidities, medication profiles, and blood test z-scores, three iterations of XGBoost classifiers were trained for broad cancer, colorectal cancer, and cardiovascular disease prediction. Model interpretability was achieved through SHAP and dimensionality reduction analyses (UMAP, t-SNE). Broad-category cancer models achieved ROC-AUC = 0.607–0.706. Colorectal cancer prediction demonstrated excellent discrimination (ROC-AUC = 0.816–0.993), and cardiovascular models showed clinical utility, notably for hypertension (ROC-AUC = 0.813, F1 = 0.861). SHAP revealed consistent importance of age, sex, basophil count, and cystatin C. Immune digital blood twins as an agnostic tool demonstrate proof-of-concept feasibility for accessible, low-cost, and scalable screening of cancer and cardiovascular diseases, supporting future integration into predictive and preventive healthcare.

* Corresponding author: hector.zenil@kcl.ac.uk

# 1. INTRODUCTION

Cancer remains one of the most prevalent non-communicable diseases worldwide. In 2022, there were 19,976,499 new cancer cases globally and 454,954 in the United Kingdom, contributing to 181,807 deaths and 1,435,322 five-year prevalent cases in the UK alone [1, 2]. Although national screening programmes exist, such as the Faecal Immunochemical Test for bowel cancer and the targeted lung health check for lung cancer [3], these cover only specific cancer types, require specialised infrastructure, and are not routinely available to the general population. The COVID-19 pandemic further disrupted screening, with 2.1 million people in the UK missing cancer tests, resulting in potential losses of 32,700 Quality-Adjusted Life Years (QALY) and increased mortality [4–7]. In 2023–24, 36 of 64 bowel-screening centres failed to meet the 90 % timeliness threshold, and regional disparities ranged from 88.8 % in London to 69.4 % in the Southwest [8]. These limitations underscore the urgent need for accessible, broad-scope, and data-driven cancer screening methods.

The complete blood count (CBC) is among the most frequently performed clinical tests [9, 10]. It measures red and white blood cell counts, platelets, haemoglobin, haematocrit, and mean corpuscular indices, reflecting both haematologic and immunologic status. Current practice interprets CBC results against broad population reference intervals, which are often too wide to detect early disease signals [11]. In contrast, intra-individual CBC variation is narrow and stable over both short [12] and long [11] timeframes, allowing personalised baselines to act as sensitive indicators of health changes. Physiological stressors such as spaceflight can transiently disturb CBC, causing spaceflight anaemia, leukopenia, and thrombocytopenia [13, 14], yet values typically return to homeostatic setpoints [11, 15]. Persistent deviations may therefore signal aging, chronic illness, or malignancy.

Cancer commonly perturbs haematologic indices. Cancer-associated anaemia impairs oxygen delivery [16]; abnormal erythrocyte morphologies such as acanthocytes [17] and stomatocytes [18] reduce deformability and microvascular flow [19]. Macrocytosis and elevated mean cell volume occur in myelodysplastic syndrome [20] and in chemotherapy-treated breast and lung cancers [21, 22]. Leukocytosis arises from paraneoplastic inflammation [23], whereas leukopenia results from marrow suppression [24]. A high lymphocyte count prior to immunotherapy correlates with improved survival [25], while an elevated neutrophil-to-lymphocyte ratio (NLR) predicts poorer prognosis in ovarian [26] and other solid tumours [27]. Platelet–leukocyte interactions and high platelet-to-lymphocyte ratios (PLR) further indicate impaired immune surveillance and adverse outcomes [28].

Recent studies have trained machine learning (ML) models to predict cancer risk from CBC and demographic data [29–33]. Most target single cancer types, e.g., gastrointestinal [30], colorectal [31, 32], or pancreatic [33], within fixed timeframes, leaving no unified framework for broader categories such as *solid*, *blood*, or *benign* cancers. This project therefore seeks to develop digital blood twins, ML-powered models integrating longitudinal blood-test trends with demographics and comorbidities for general cancer and cardiovascular disease screening. By enabling early detection through routine data, such tools could reduce healthcare disparities and save an estimated £210 million annually if screening availability in remote regions matched that of urban centres [34].

Accordingly, the central objectives of this study were to develop *digital blood twins* using state-of-the-art machine learning models for cancer and cardiovascular disease screening, to explore longitudinal blood-test dynamics and identify key predictive features for disease detection, and to evaluate the potential of these models to enable large-scale, cost-effective, and equitable population screening.

## 2. METHODS

### Data Source and Pre-processing: United Kingdom Biobank (UK Biobank)

The UK Biobank is a large prospective cohort including ~500,000 participants aged 40–69 at recruitment (2006–2010) [36, 37]. The dataset used contained blood tests, demographics, prescribed medications, comorbidities, and disease outcomes for 373,269 subjects, each with up to three blood-test timepoints, though CBC data were incomplete for later visits.
CBC and biochemical features were supplemented by derived inflammatory ratios known to associate with cancer and CVD, including the neutrophil-to-lymphocyte ratio (NLR) [26–28, 38], lymphocyte-to-monocyte ratio (LMR) [39, 40], and platelet-to-lymphocyte ratio (PLR) [28]. Blood tests were grouped into *cardiovascular*, *liver*, *renal*, *diabetes*, *bone/joint*, *cancer*, and *CBC* panels per the UK Biobank biomarker classification [41].

Medication data included 3,741 unique drugs; only those prescribed to > 50 patients were retained to improve statistical reliability [42]. Drugs were classified using the British National Formulary (BNF 89, March 2025) [43] by paragraph-level pharmacologic grouping, and prescriptions at each timepoint were one-hot encoded.

Cardiovascular outcomes included angina, myocardial infarction, hypertension, and stroke, each one-hot encoded. Cancer diagnoses were categorised by timing (*prevalent* if before the first blood test, *incident* if after) and by type: solid (C00–C80), blood (C81–C96), or benign/in situ (D codes). Subjects lacking ICD-10 codes (n = 3,498) were excluded. The final dataset contained six one-hot-encoded classes: prevalent and incident forms of solid, blood, and benign cancers.

### Exploratory Data Analysis

Incidences of solid, blood, and benign cancers were computed overall and by sex, alongside cardiovascular disease prevalence. The most commonly prescribed medications were ranked by prescription frequency. Using normalized values stratified by sex and age, mean ± SD of each biomarker were calculated for cancer, CVD, and control groups. Point-biserial correlation heatmaps (Figure 3) visualised associations between blood panels and disease categories, revealing multi-systemic biomarker signatures relevant to cancer and cardiovascular risk.

### Machine Learning Model

#### XGBoost.

Extreme Gradient Boosting (XGBoost) is an advanced ensemble learning algorithm that builds multiple decision trees sequentially using gradient descent to minimize classification loss [44]. It has demonstrated high performance in early lung cancer prediction using CBC and clinical data [45]. In this study, XGBoost was employed with a binary logistic loss function to classify cancer status. Both L1 (Lasso) and L2 (Ridge) regularization were incorporated to prevent overfitting by penalizing large model weights, while second order (Hessian-based) optimization enabled faster convergence and efficient handling of large-scale data.

Given substantial missingness in blood test results, especially at timepoints 1 and 2, XGBoost's native capability to handle missing data was utilized. Missing entries were treated as a distinct category, and the algorithm dynamically determined optimal assignment during node splitting based on log-loss reduction, eliminating the need for potentially biased imputation.

The model was implemented using the Python xgboost library, with datasets formatted into DMatrix objects to automatically recognise NaN values. Hyperparameters including learning rate, maximum tree depth, and number of boosting rounds were optimized via randomized search cross-validation. Regularization parameters *alpha* (L1) and *lambda* (L2) were fine-tuned to balance bias and variance. To address class imbalance, the scale_pos_weight parameter was derived from the ratio of negative to positive cases, increasing the penalty for misclassifying minority cancer classes.

**Machine Learning Model Evaluation**

The machine learning models were evaluated based on multiple metrics to assess the predictive accuracy, especially when the dataset has a severe class imbalance. The following metrics were used for model evaluation:

1. Receiver Operating Characteristic – Area Under the Curve (ROC-AUC): it plots the recall (true positive rate) against false positive rate, with a higher value closer to 1 indicating better discriminative power. It measures the model's ability to differentiate positive and negative classes.

2. Precision-Recall – Area Under the Curve (PR-AUC): it measures the tradeoff between precision and recall

3. Threshold: the optimal decision threshold used to convert probability scores into binary predictions of cancer type, selected through maximizing the F1 score. It has values ranging from 0 to 1.

4. Precision: The proportion of predicted positive cancer cases that are correct, higher precision indicates fewer false positives. It has values ranging from 0 to 1.
$$\text{Precision} = \frac{True\ Positives}{True\ Positives + False\ Positives}$$

5. Recall (sensitivity): The proportion of actual positive cases correctly identified, higher recall indicates fewer missed positive cases. It has values ranging from 0 to 1.
$$\text{Recall} = \frac{True\ Positives}{True\ Positives + False\ Negatives}$$

6. Specificity: The proportion of actual negative cases correctly identified, higher specificity indicates fewer false positives among negative cases. It has values ranging from 0 to 1.
$$\text{Specificity} = \frac{True\ Negative}{True\ Negatives + False\ Positives}$$

7. F1 score: The harmonic mean of precision and recall (sensitivity). It balances precision and recall, making it suitable for imbalanced datasets. Higher F1 score indicate better overall performance for imbalanced datasets.

$$\text{F1 score} = \frac{2 \times precision \times recall}{precision + recall}$$

8. Additional diagnostic metrics. To complement our performance metrics, we computed Balanced Accuracy ($\frac{\text{sensitivity+specificity}}{2}$), Likelihood Ratios (LR+ = sensitivity / [1 − specificity]; LR− = [1 − sensitivity] / specificity), and the Diagnostic Odds Ratio (DOR) = LR+/LR− at the chosen operating thresholds.

**Feature Importance Analysis with SHAP**

Feature importance was interpreted using SHapley Additive exPlanations (SHAP), a game-theoretic framework that quantifies each feature's contribution to the model's predicted probability [46]. For each XGBoost cancer model, SHAP values were computed and exported to CSV files, and the top 10 features by mean absolute SHAP value were visualized in SHAP summary plots.

In these plots, the x-axis represents the SHAP value, where positive values indicate features driving predictions toward cancer, and negative values toward control. The y-axis lists features ranked by overall importance. Each point corresponds to a single subject, coloured by the feature's raw value (blue = low, red = high). The distance from zero denotes the strength of influence.

SHAP summary plots were generated for all three XGBoost model iterations, broad cancer, colorectal cancer, and cardiovascular disease, revealing globally consistent predictors such as age, sex, basophil count, and cystatin C.

| Model | Blood tests | Prescribed medications | Comorbidities | Sex | Age | Control group |
|---|---|---|---|---|---|---|
| 1 | ✓ (Standard scaler) | ✓ | | | | No cardiovascular diseases and cancer + all blood tests within reference interval |
| 2 | ✓ (Standard scaler) | | ✓ | ✓ | | No cardiovascular diseases and cancer + all blood tests within reference interval + no other comorbidities which may affect blood test results |
| Final | ✓ (z-score based on sex and age) | ✓ | ✓ | ✓ | ✓ | |

**Table 1** Data incorporated into 3 iterations of XGBoost models

**Dimensionality Reduction and Machine Learning Models**

To visualize high-dimensional relationships and disease separability, Uniform Manifold Approximation and Projection (UMAP) and t-Distributed Stochastic Neighbor Embedding (t-SNE) were applied to training data. UMAP (*umap-learn*, random state = 42) reduced features to two components by minimizing cross-entropy between high- and low-dimensional neighbor graphs [47]. t-SNE (*sklearn.manifold.TSNE*, perplexity = 30, random state = 42) minimized divergence between pairwise similarity distributions using a Student-t kernel [48]. Resulting embeddings were visualized with *matplotlib* and *seaborn*, coloured by disease group.

For model construction, XGBoost classifiers were trained to predict six cancer outcomes (prevalent/incident solid, blood, and benign cancers). Blood tests were standardized, sex was one-hot encoded, and medications were encoded per timepoint. The control group comprised 204,504 subjects without cancer or cardiovascular disease; those with out-of-range biomarkers were excluded using UK Biobank and literature-defined reference intervals [49, 50], yielding 14,794 controls. Missing values were handled natively by XGBoost. Data were split 80 / 20 using stratified sampling (random state = 42).

Separate binary models were trained for each cancer class with initial parameters: 2,000 estimators, learning rate = 0.03, max depth = 5, subsample = 0.8, and L2 = 1.0. Class imbalance was addressed via scale_pos_weight, and thresholds were tuned for maximal F1. RandomizedSearchCV (3-fold) optimized hyperparameters (learning rate 0.01–0.3, depth 3–9, estimators 100–1000).

Because prescribed medications contributed minimal predictive value (low SHAP scores), they were excluded in a second model incorporating comorbidities and sex, leaving 8,825 clean controls. Comorbidities were one-hot encoded by ICD-10 (≥ 100 cases) after excluding confounding conditions. Only baseline (timepoint 0) blood tests were retained to reduce missingness.

The final model integrated prescribed medications, comorbidities, age, and sex. To account for biological variation, all blood-test features were z-scored by age and sex [51, 52]. Optimized models were retrained with early stopping and threshold tuning, and feature importance was interpreted using SHAP. The z-score of each blood test parameter was calculated with the following formula:

$$z = \frac{x - \mu}{\sigma}$$

Z = z-score stratified by sex and age
x = raw value of blood test parameter
$\mu$ = mean stratified by sex and age
$\sigma$ = standard deviation stratified by sex and age

Z-scores were used instead of the raw values or standardized blood test results for the machine learning model.

**Final, Colorectal, and Cardiovascular Models**

The final digital blood-twin model integrated age, sex, prescribed medications, comorbidities, and blood-test z-scores for cancer prediction. Data were split 80 / 20 using stratified sampling (*train_test_split*, random state = 42). Because timepoints 1–2 contained sparse data, two XGBoost models (with and without them) were compared; similar performance justified retaining all timepoints to support future longitudinal applications. Class imbalance was addressed with scale_pos_weight, and default parameters (learning rate = 0.3, max depth = 6) were used with early stopping (patience = 100, *auprc* metric). Control subjects lacked cancer, cardiovascular, or confounding comorbidities, consistent with earlier models. Hyperparameters were tuned via RandomizedSearchCV (3-fold CV, 20 combinations) for learning rate 0.01–0.3, max depth 3–9, and estimators 100–1000; optimized models underwent threshold tuning, SHAP interpretation, and dimensionality-reduction visualization.

For colorectal cancer, one of the most common UK Biobank cancers [31, 32], age, sex, medications, and z-scored blood tests were used as predictors. Cases were defined by ICD-10 C18–C20, classified as 6178 total, with 887 prevalent and 822 incident. Three separate models predicted overall, prevalent, and incident disease. Data splitting, weighting, and tuning followed the cancer pipeline, using 15 hyperparameter combinations (learning rate 0.01–0.3, depth 5–9, estimators 500–1000).

For cardiovascular disease, subjects with ICD-10 “I” codes were excluded from controls and predictors. Four binary XGBoost classifiers predicted angina (n = 12,161), heart attack (8870), hypertension (102,561), and stroke (5939) using age, sex, medications, and age/sex-based z-scores. Parameter tuning matched the colorectal model. Performance was assessed via ROC-AUC, PR-AUC, precision, recall, specificity, and F1, with SHAP and embedding analyses confirming feature interpretability.

# 3. RESULTS

## Cancer incidence

The total incidences of solid, blood and benign cancer were first calculated. Then, the cancers were subclassified into incident and prevalent with their incidences calculated. A total of 82,826 patients suffered from cancer. There were 70,098 cases of solid cancer, 4,905 cases of blood cancer, 14,037 cases of benign cancer (Figure 3A). Solid cancer was the most common category of cancer, with 25,216 prevalent cases and 48,527 incident cases. Benign cancer came second in the number of incidents, with 5,397 prevalent cases and 8,017 incident cases. Blood cancer was the least common, with only 1,200 prevalent cases and 3,556 incident cases. Prevalent versus incident stratification across cancer types is shown **(Figure 3B)**

The most common cancer recorded in the UK biobank database was Non-Melanoma Skin Cancer (26,944 cases), followed by Breast Cancer (14036) and Prostate Cancer (10689). Top 10 most common cancers in both sexes are visualized **(Figure 3C)**. The most prevalent cancer in male was Non-Melanoma Skin Cancer (14360), followed by Prostate Cancer (10688) and Colon Cancer (2208). The distribution for male cancers is shown **(Figure 3D).** The most prevalent cancer in female was Breast Cancer (13956), followed by Non-Melanoma

Skin Cancer (12584) and Melanoma (2048). The analogous female-specific distribution is shown **(Figure 3E)**.

## Cardiovascular disease incidence

A majority of the patients (260439 patients) did not suffer from any cardiovascular diseases. There were 12161 (4005 female, 8156 male) cases of angina, 8870 (1748 female, 7122 male) cases of heart attack, 102561 (49736 female, 52825 male) cases of hypertension and 5939 (2447 female, 3492 male) cases of stroke. Sex-stratified cardiovascular disease incidences are visualized (Figure 3F). he relative sex distribution of hypertension cases is detailed (Figure 3G)

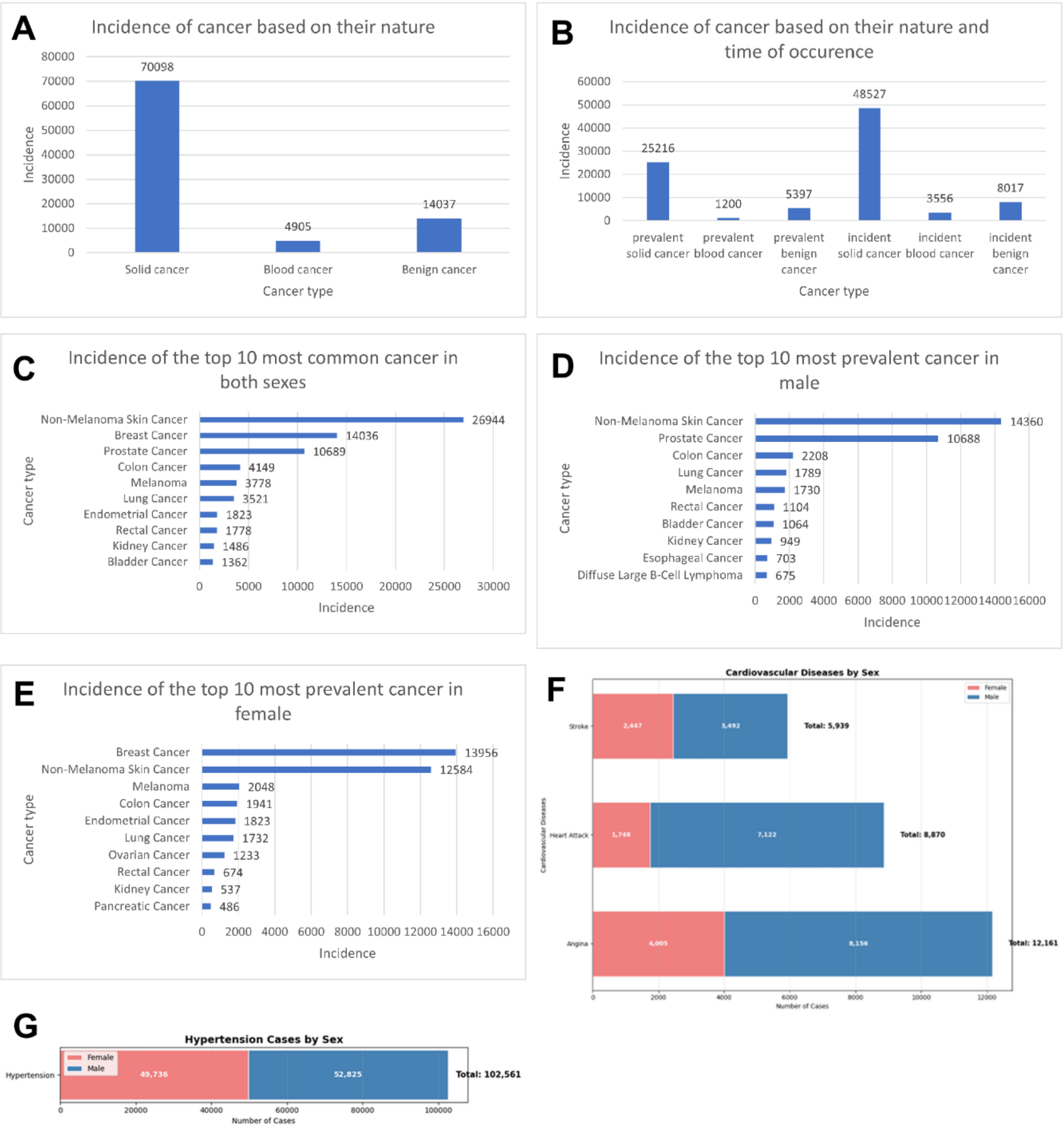

**Figure 3.** A) Incidence of cancer based on their nature. B) Incidence of cancer based on their nature and time of occurrence. C) Incidence of the top 10 most common cancer in both sexes. D) Incidence of the top 10 most common cancer in male. E) Incidence of the top 10 most common cancer in female. F) Incidence of the stroke, heart attack and angina by sex. G) Incidence of hypertension by sex.

## Cancer Correlation Heatmaps

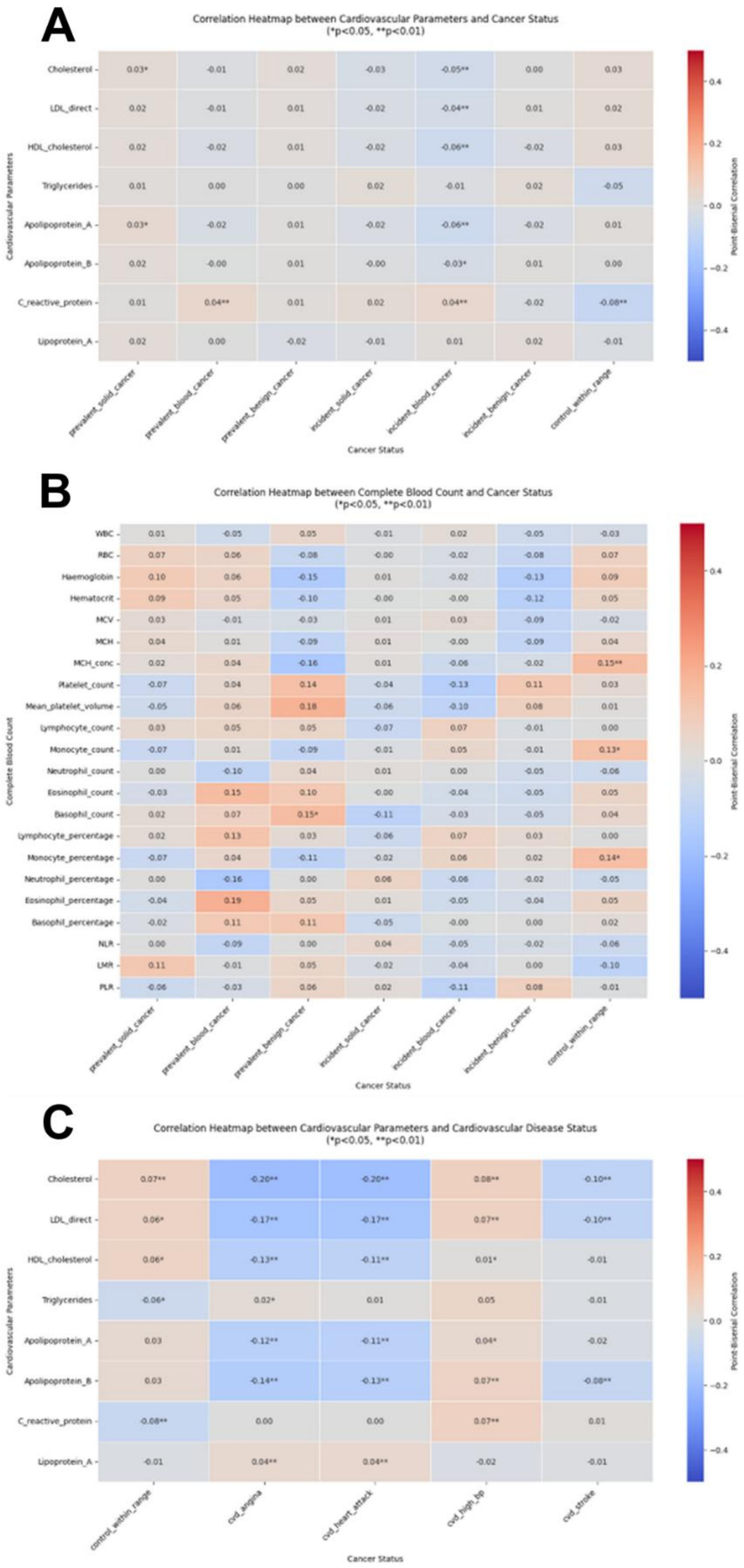

**Figure 4. A)** Correlation Heatmap between Cardiovascular Parameters and Cancer Status. **B)** Correlation Heatmap between CBC and Cancer Status. **C)** Correlation Heatmap between Cardiovascular Parameters and Cancer Status.

As observed from the correlation heatmaps, there was not a single blood test parameter with a magnitude of point-biserial correlation coefficient > 0.2 and p-value >0.05. Basophil count was weakly positively correlated with prevalent benign cancer (point-biserial correlation coefficient +0.15, p-value<0.05) (Figure 4B). Hence, there were no statistically significant correlations between single blood test parameters and cancer statuses. [53] Combining multiple blood test parameters was necessary for meaningful cancer prediction (Figure 4A).

## Cardiovascular disease Correlation Heatmaps

Cardiovascular parameters, which consist of lipid profile were more correlated with cardiovascular disease statuses compared to other parameters (Figure 4C). Cholesterol level was weakly negatively correlated with angina (point-biserial correlation coefficient -0.20, p-value <0.01), heart attack (point-biserial correlation coefficient -0.20, p-value <0.01) and stroke (point-biserial correlation coefficient -0.10, p-value <0.01). Similarly, LDL direct level was weakly negatively correlated with angina (point-biserial correlation coefficient -0.17, p-value <0.01), heart attack (point-biserial correlation coefficient -0.17, p-value <0.01) and stroke (point-biserial correlation coefficient -0.10, p-value <0.01). HDL cholesterol level was weakly negatively correlated with angina (point-biserial correlation coefficient -0.13, p-value <0.01) and heart attack (point-biserial correlation coefficient -0.11, p-value <0.01). Apolipoprotein A level was weakly negatively correlated with angina (point-biserial correlation coefficient -0.12, p-value <0.01) and heart attack (point-biserial correlation coefficient -0.11, p-value <0.01). Apolipoprotein B level was weakly negatively correlated with angina (point-biserial correlation coefficient -0.14, p-value <0.01) and heart attack (point-biserial correlation coefficient -0.13, p-value <0.01)

Although there were more significant correlations between blood test parameters and cardiovascular diseases, particularly in lipid profile, there were still no blood test parameters with magnitude of point-biserial correlation coefficient > 0.2 and p-value >0.05. Hence, there were no statistically significant correlations between single blood test parameters and cancer statuses. [53] Thus, combining multiple blood test parameters was necessary for meaningful cardiovascular disease prediction.

## *Model 1: incorporating prescribed medications*

The evaluation metrics for each cancer outcome on the test set of both untuned and tuned XGBoost models are summarized below. Since SHAP summary plots showed consistent feature importance patterns across all cancer models, only the final model's plots are presented. Model performance varied across the six cancer outcomes, with rarer cancers showing poorer predictive accuracy due to class imbalance. Hyperparameter tuning generally improved performance, particularly for low-prevalence cancers.

The models demonstrated modest ability to distinguish cancer from control cases based on ROC-AUC values (Figure 5A). In the untuned models, ROC-AUC ranged from 0.5420 for incident benign cancer to 0.6774 for incident solid cancer. After tuning, values improved for prevalent blood cancer from 0.5695 to 0.6970 and for prevalent benign cancer from 0.6372 to 0.6911, while incident solid cancer remained nearly unchanged at 0.6770 since optimization prioritized the F1 score rather than ROC-AUC.

Threshold optimization lowered the probability cut-offs, increasing recall at the expense of specificity, which is appropriate for imbalanced datasets emphasizing detection of positive cases. Precision increased slightly after tuning, rising from 0.0582 to 0.1458 in prevalent blood cancer prediction, reflecting fewer false positives and improved reliability of positive classifications. Recall also improved for less common cancers, including prevalent blood cancer (0.0466 to 0.1186), prevalent benign cancer (0.3573 to 0.4159), and incident blood cancer (0.1084 to 0.1700), but decreased slightly for prevalent solid cancer (0.8074 to 0.7653) and incident solid cancer (0.9316 to 0.8991), indicating a trade-off to enhance specificity and F1 score.

Specificity increased for most cancer predictions, particularly for prevalent solid cancer (0.3652 to 0.4167), incident solid cancer (0.2834 to 0.3536), and incident benign cancer (0.5462 to 0.5806), demonstrating improved identification of true negative cases. F1 score, the harmonic mean of precision and recall, improved in most cancer classifications, such as prevalent blood cancer (0.0518 to 0.1308), prevalent benign cancer (0.1549 to 0.1776), and incident solid cancer (0.7029 to 0.7058). Thus, hyperparameter tuning substantially enhanced the performance and balance of the XGBoost models, particularly improving prediction of rare cancer types.

## *Model 2: incorporating comorbidities and sex*

The evaluation metrics for each cancer outcome on the test set of the tuned XGBoost models are shown for comparison with the previous iteration. Since SHAP summary plots revealed consistent feature importance patterns across all models, only the final model's plots are presented. As with the first model, predictive ability varied across cancer outcomes. Due to dataset imbalance, rarer cancers showed poorer performance, while hyperparameter tuning generally improved results, particularly for less prevalent types.

The ROC-AUC ranged from 0.6353 for prevalent solid cancer to 0.7072 for prevalent benign cancer, underscoring the importance of incorporating comorbidities in cancer prediction, especially for blood cancers (see Supplementary Tables). Precision values ranged from 0.1128 for prevalent blood cancer to 0.5834 for incident solid cancer, while recall ranged from 0.1602 to 0.9312 for the same categories. Both precision and recall were comparable to those from the first model, with solid cancer maintaining the highest performance across metrics.

Specificity ranged from 0.2518 for incident solid cancer to 0.9839 for prevalent blood cancer, and F1 scores ranged from 0.1324 to 0.7174, indicating balanced improvements. Notably, incident blood cancer achieved a marked increase in F1 score, from 0.1273 in the first model to 0.2488 in the second. Although prevalent blood, benign, and incident blood cancer models achieved higher ROC-AUC values, their sensitivities remained low, reflecting a tendency to classify positive cases as controls. In contrast, incident solid cancer, despite moderate AUC, achieved a strong F1 score of 0.7174, indicating greater sensitivity and fewer false positives.

## *Final model incorporating prescribed medications, comorbidities, sex and age*

The evaluation metrics computed for each cancer outcome on the test set of the tuned XGBoost models shown for comparison with the previous tuned model. The SHAP summary plots and dimensionality reduction analysis were also included for the final model.

Similar to the first two models, the final model had varying ability to predict different cancer outcomes. Due to the imbalanced dataset, rarer cancer types suffered from poorer prediction performance. Hyperparameter tuning generally improved model performance, especially for less prevalent cancer types.

After hyperparameter tuning, the ROC-AUC ranged from 0.6070 (incident benign cancer) to 0.7062 (prevalent blood cancer), which were similar to that of the second iteration of models. The PR-AUC and thresholds were similar to the second iteration of models.
Recall ranged from 0.1299 (prevalent blood cancer) to 0.8251 (prevalent solid cancer). Recall of some cancer types, such as prevalent solid cancer (from 0.7303 to 0.8251), prevalent benign cancer (from 0.3396 to 0.3858) and incident benign cancer (from 0.4452 to 0.4774) improved compared to the previous iteration of models. But there were slight drops in recalls of the other cancer prediction models.

Specificity ranged from 0.2316 (incident solid cancer) to 0.991 (prevalent blood cancer), showing varied ability in picking up control cases, with particularly weak ability in identifying control cases from solid cancer cases. F1 score ranged from 0.1467 (prevalent blood cancer) to 0.7131 (incident solid cancer), which were similar to the results of the second iteration of models. At the selected thresholds, Incident Solid Cancer achieved a Balanced Accuracy = 0.585, with $LR^+ = 1.23$, $LR^- = 0.067$, and DOR = 18.3, indicating strong rule-out utility despite modest specificity. Incident Blood Cancer and Prevalent Blood Cancer showed DOR = 8.16 and 16.4, respectively, supporting clinically meaningful discrimination in rare classes.

| **Cancer Type** | **ROC-AUC** | **PR-AUC** | **Threshold** | **Precision** | **Recall** | **Specificity** | **F1** |
|---|---|---|---|---|---|---|---|
| Prevalent Solid Cancer | 0.6389 | 0.3969 | 0.439 | 0.3272 | 0.8251 | 0.3461 | 0.4686 |
| Prevalent Blood Cancer | 0.6970 | 0.0753 | 0.693 | 0.1684 | 0.1299 | 0.9910 | 0.1467 |
| Prevalent Benign Cancer | 0.7062 | 0.1400 | 0.551 | 0.1325 | 0.3858 | 0.8478 | 0.1973 |
| Incident Solid Cancer | 0.6665 | 0.6580 | 0.287 | 0.5754 | 0.9373 | 0.2316 | 0.7131 |
| Incident Blood Cancer | 0.6993 | 0.1442 | 0.573 | 0.1931 | 0.2830 | 0.9538 | 0.2295 |
| Incident Benign Cancer | 0.6070 | 0.1330 | 0.500 | 0.1256 | 0.4774 | 0.6793 | 0.1989 |

**Table 2.** Per-label XGBoost model (final model) performance after hyperparameter tuning

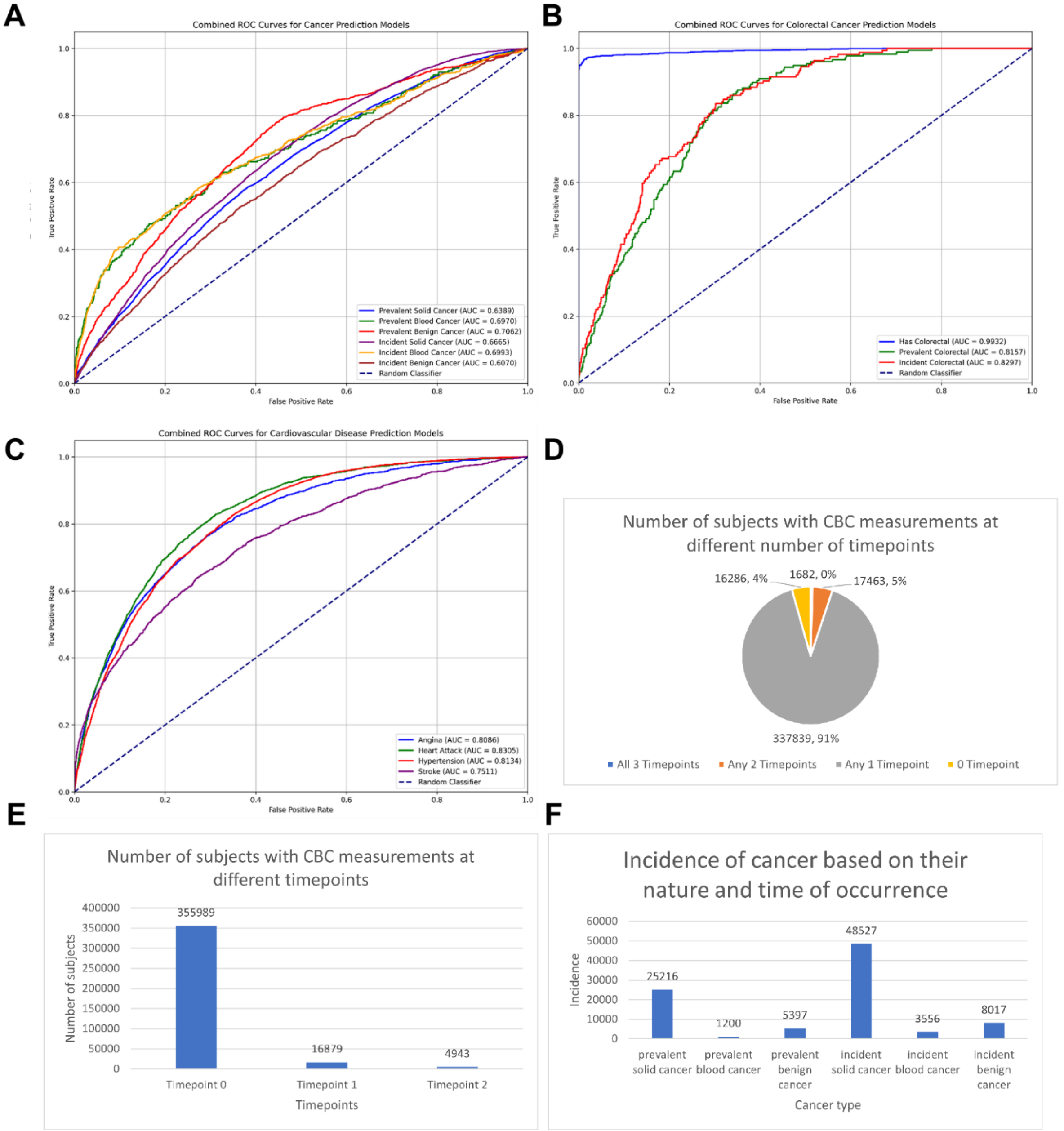

**Figure 5. A)** Receiver Operating Characteristic (ROC) Curves for final cancer prediction models. B) ROC Curves for colorectal cancer prediction models. C) ROC Curves for cardiovascular disease prediction models. D) Number of subjects with CBC measurements at different number of timepoints. E) Number of subjects with CBC measurements at different timepoints. F) Incidence of cancer based on their nature and time of occurrence

## Feature Importance Analysis

The top 10 features with the highest mean absolute SHAP values, which indicate feature importance in model training, were plotted for each cancer prediction model respectively. Sex and age remained important factors in most cancer prediction. For both prevalent and incident solid cancer, older age was more predictive of solid cancer diagnosis.

For CBC metrics, basophil count was particularly significant in solid and benign cancer diagnosis. A lower basophil count was more indicative of solid or benign cancer diagnosis. Cystatin C is another top feature in most cancer classifications. Higher cystatin C level was more suggestive of prevalent cancers and incident blood cancers (See Table S8 in Supplementary Information).

### *Model for colorectal cancer prediction*

| Group Type | ROC-AUC | PR-AUC | Precision | Recall | Specificity | F1 Score |
|---|---|---|---|---|---|---|
| Has Colorectal Cancer | 0.9932 | 0.9933 | 0.9907 | 0.9523 | 0.9938 | 0.9711 |
| Prevalent Colorectal Cancer | 0.8157 | 0.1927 | 0.1586 | 0.6328 | 0.7897 | 0.2537 |
| Incident Colorectal Cancer | 0.8297 | 0.2229 | 0.1751 | 0.6524 | 0.8223 | 0.2761 |

**Table 3.** Per-label XGBoost model (colorectal cancer) performance after hyperparameter tuning

The XGBoost model demonstrated significantly better performance in predicting colorectal cancer compared with broader cancer categories such as solid, blood, and benign cancers. The ROC-AUC values for prevalent (0.8157) and incident colorectal cancer (0.8297) exceeded the 0.8 threshold for clinical utility [53] and were higher than those achieved by broad-category models. The "Has Colorectal Cancer" model achieved a Balanced Accuracy = 0.973, $LR^{+} \approx 154$, $LR^{-} \approx 0.048$, and DOR $\approx 3.22 \times 10^{3}$, consistent with excellent diagnostic power. The prevalent and incident subtypes also showed strong discrimination, with DOR values of 6.49 and 8.68, respectively, approaching clinical-grade thresholds for confirmatory testing.

Colorectal cancer prediction demonstrated robust discriminative performance across both subtypes (Figure 5B). The overall model achieved ROC-AUC = 0.9932 and F1 = 0.9711, reflecting high accuracy in identifying colorectal cancer cases. However, it was less effective at distinguishing between incident and prevalent cancers, likely due to overlapping features and limited sample sizes for each subtype. The high ROC-AUC values were primarily driven by high specificity (0.7897 for prevalent and 0.8223 for incident colorectal cancer). In contrast, recall remained moderate (0.6328 and 0.6524, respectively), indicating lower sensitivity for detecting positive cases, while low precision (0.1586 and 0.1751) reflected a relatively high false-positive rate.

Feature importance analysis identified older age and comorbidity with hypertension as key predictors of colorectal cancer. Consistent with previous broad cancer models, low basophil count emerged as a significant feature, suggesting its potential as a general hematologic biomarker. Additionally, higher mean corpuscular volume (MCV) and mean corpuscular hemoglobin (MCH) were associated with incident colorectal cancer, while elevated cystatin C levels were linked to prevalent colorectal cancer.

### *Model for prediction of cardiovascular diseases*

| Group Type | ROC-AUC | PR-AUC | Precision | Recall | Specificity | F1 Score |
|---|---|---|---|---|---|---|
| Angina | 0.8086 | 0.3410 | 0.2342 | 0.7336 | 0.7336 | 0.3551 |

| Group Type | ROC-AUC | PR-AUC | Precision | Recall | Specificity | F1 Score |
|---|---|---|---|---|---|---|
| Heart Attack | 0.8305 | 0.2863 | 0.1918 | 0.7661 | 0.7460 | 0.3068 |
| Hypertension (high bp) | 0.8134 | 0.9498 | 0.9296 | 0.8018 | 0.6733 | 0.8610 |
| Stroke | 0.7511 | 0.2210 | 0.1087 | 0.6305 | 0.7346 | 0.1855 |

**Table 4.** Per-label XGBoost model (cardiovascular disease) performance after hyperparameter tuning

The ROC-AUC in cardiovascular disease predictions were significantly higher than in broad category cancer prediction. It ranged from 0.7511 (stroke) to 0.8305 (heart attack).

The model for predicting hypertension had the best overall performance. It had the highest sensitivity in predicting positive cases (0.8018), and a high precision (0.9296), indicating little false positives. The specificity (0.6733) was acceptable. It was likely due to the significantly higher incidence of hypertension in the dataset.

The other models had satisfactory recall, ranging from 0.6306 (stroke) to 0.7661 (heart attack), indicating modest ability in detecting positive cases. This was at the expense of low precision, ranging from 0.1087 (stroke) to 0.2342 (angina). This indicated that the models classified more false positive cases to pick up more positive cases. Cardiovascular models yielded robust class-balanced performance (e.g., Heart Attack: Balanced accuracy = 0.756, $LR^{+}$ = 3.02, $LR^{-}$ = 0.314, DOR = 9.63; Hypertension: Balanced accuracy (BA)= 0.738, DOR = 8.35), supporting potential clinical screening and translational utility for these outcomes.

From the SHAP summary plots of the four cardiovascular disease predictions, age and sex were important factors for prediction. Advanced age was highly predictive of all four cardiovascular diseases. In terms of comorbidities, pain in the throat and chest (ICD-10 code R07) was highly suggestive of angina and heart attack. Blood biochemistry test results played a significant role in predicting cardiovascular disease. Cholesterol levels, direct LDL levels and HDL levels were associated with cardiovascular disease risk, concurrent with its impact on increasing risk of atherosclerosis. Higher glycated haemoglobin (HbA1c) levels were also predictive of angina, heart attack and stroke. Basophil count was the only CBC metric among the top features in predicting cardiovascular diseases. A lower basophil count was highly suggestive of hypertension.

## 4. DISCUSSION

Cancer prediction models showed that blood cancer classifiers achieved the highest ROC-AUC values, reflecting a stronger ability to differentiate blood cancer from controls, consistent with evidence that abnormal blood tests more readily indicate hematologic malignancies [54, 55]. However, their F1 scores, sensitivity, and precision were lower than for solid cancers due to limited training data. Solid cancer models displayed higher sensitivity and F1 scores but lower specificity, likely caused by lower probability thresholds that led to over-prediction of cancer among controls. For blood and benign cancers, higher specificity contributed to their ROC-AUC, yet class imbalance caused frequent misclassification as controls, reducing sensitivity and precision.

Incident cancer models demonstrated modest improvement across iterations, with the incident blood cancer model rising from a ROC-AUC of 0.6262 to 0.6993 and similar gains in incident benign cancers, emphasizing the value of incorporating comorbidities. Nonetheless, all cancer models remained below the 0.8 AUC threshold for clinical utility [56], likely due to limited longitudinal data, broad cancer categories, and class imbalance. Dimensionality reduction revealed overlapping feature projections between cancers and controls (Supplementary Information), confirming that blood test results alone only weakly correlated with disease status. Features such as low basophil count emerged consistently across models, suggesting potential as a biomarker for general cancer screening.

Confounding factors, including race and undiagnosed comorbidities, may also contribute to inter-individual blood test variation despite age- and sex-standardized z-scores. Broad categories like "solid" or "blood" cancer encompass distinct diseases with overlapping hematologic profiles; for example, colorectal cancer reduces hemoglobin via chronic rectal bleeding [57], while multiple myeloma decreases red blood cell production through marrow infiltration [58]. When models targeted specific diagnoses, such as colorectal cancer, predictive performance improved markedly due to clearer feature separation, highlighting the potential of multiple specialized models for distinct cancer types rather than a single broad classifier.

Cardiovascular disease models performed substantially better than cancer models. Classifiers for angina, heart attack, and hypertension all achieved ROC-AUC values above 0.8 (Figure 5C), indicating clinical-grade discrimination. This improvement likely reflects stronger correlations between cardiovascular outcomes and biochemical parameters such as lipid profiles. Hypertension prediction achieved particularly high precision and recall due to large sample size, while smaller case numbers for angina, myocardial infarction, and stroke reduced their precision. These findings demonstrate the digital blood twin's applicability to other non-communicable diseases.

The SOMA dataset's small sample size (n = 4) limited statistical power, and CBC data were not collected during spaceflight, preventing capture of immediate hematological shifts. In the UK Biobank, missing data were extensive, where 91 % of subjects had CBC from only one timepoint, and just 1,682 (0.0045 %) had three longitudinal tests. The absence of diagnosis dates in several cancer cases hindered clear classification into incident versus prevalent disease. Moreover, severe class imbalance favored solid cancers (prevalent 25,216; incident 48,527) over blood cancers (prevalent 1,200; incident 3,556) (Figure 5F), lowering minority-class sensitivity. The broad categorization of cancers also obscured organ-specific signatures [59], suggesting that future models should stratify by organ system.

**Limitations and Prospects**

Despite these limitations, the project establishes a proof-of-concept for digital blood twins as low-cost, accessible screening tools. Their short training time, modest computational requirements, and open-source framework lower barriers to adoption. By enabling early detection of cancer and cardiovascular diseases, especially in remote areas, this approach could potentially save the NHS £210 million annually [34] and reduce disparities in healthcare access. Regulatory approval will require compliance with MHRA and FDA standards for Software as a Medical Device (SaMD) [60, 61], likely classified as Class II (decision-support) or Class III (diagnostic) [62], along with strict adherence to GDPR for patient data protection.

Current population-level CBC reference intervals are broad and insensitive to individual variation. In contrast, intra-individual hematologic variability is stable over time [11]. Modeling personalized blood test setpoints could allow deviations to flag disease more accurately. By incorporating demographic and clinical factors, digital blood twins advance precision medicine through individualized screening. However, the UK Biobank's limited demographic range (ages 40–69 at recruitment) may restrict generalizability, necessitating inclusion of broader and more diverse datasets.

**Conclusion**

The study demonstrates that routinely collected blood test data, demographics, comorbidities, and medications can support development of digital blood twins for early disease screening. CBC stability in SOMA astronauts supports its role as a longitudinal biomarker, while UK Biobank-trained XGBoost models confirmed feasibility. Broad cancer categories yielded modest performance, but colorectal and cardiovascular models achieved clinically relevant discrimination. Feature importance analyses consistently highlighted age, sex, basophil count, and cystatin C as top predictors. Beyond AUC and F1, balanced accuracy, likelihood ratios, and diagnostic odds ratios confirmed class-balanced discrimination, particularly for colorectal and cardiovascular endpoints. CBC percentage changes post-spaceflight further validated small, physiologically consistent adaptive shifts.

Low sensitivity in rare cancers risks under-diagnosis, while low precision could lead to false positives and patient anxiety. Model generalizability is currently constrained by limited demographics and absence of multi-cohort validation. Integrating longitudinal datasets and external validation will mitigate these risks. Commercially, the model's rapid training and minimal preprocessing are advantages, but lack of long-term data remains a key barrier to scalability and regulatory readiness.

Future work will incorporate multi-year blood test trajectories to establish individual hematological setpoints and apply deep learning architectures such as multilayer perceptrons for richer pattern recognition. Specialized cancer-type models and inclusion of omics, racial, and socioeconomic features will enhance accuracy and equity. Extending digital blood twins to predict other chronic diseases will further expand their potential for precision preventive medicine.

## SUPPLEMENTARY INFORMATION

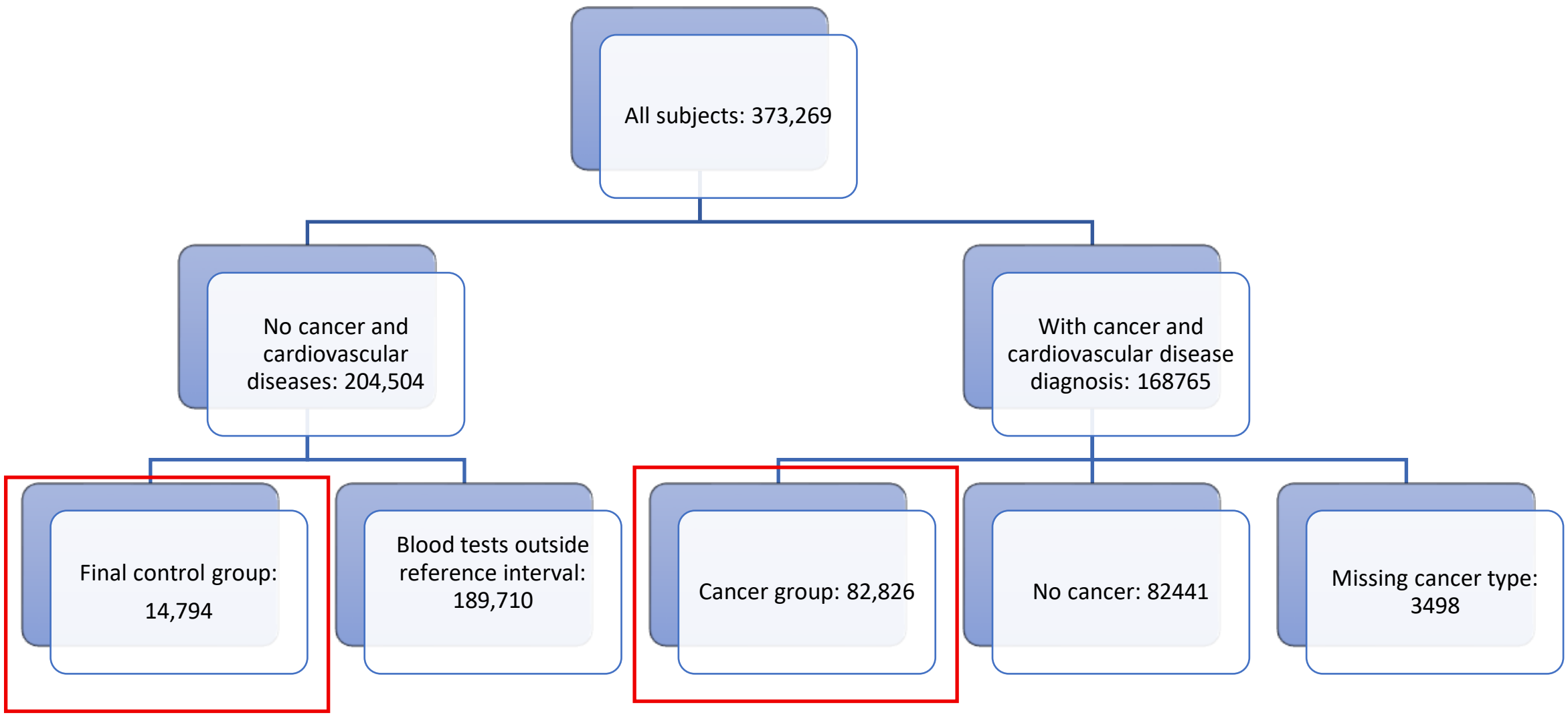

**Figure 2.2** Flowchart of control group and cancer group classification

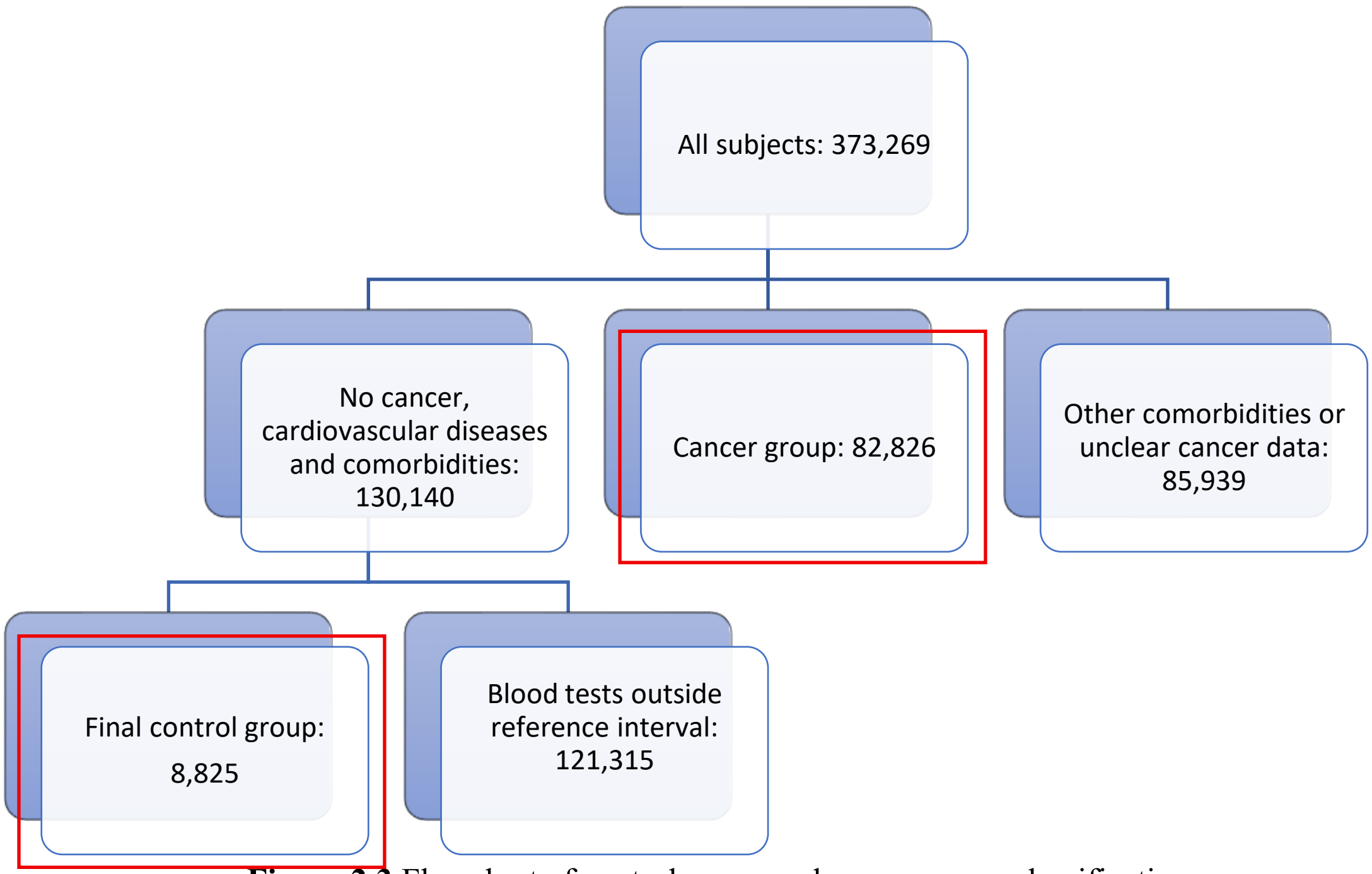

**Figure 2.3** Flowchart of control group and cancer group classification

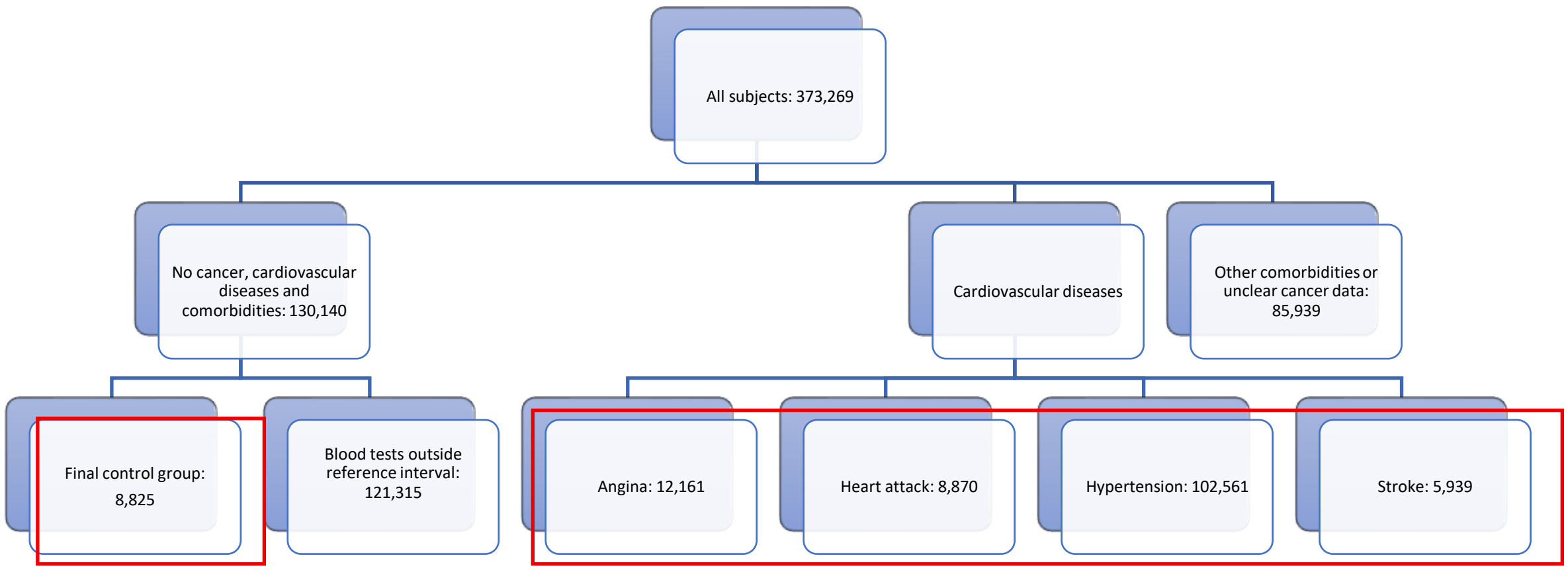

**Figure 2.4** Flowchart of control group and cancer group classification

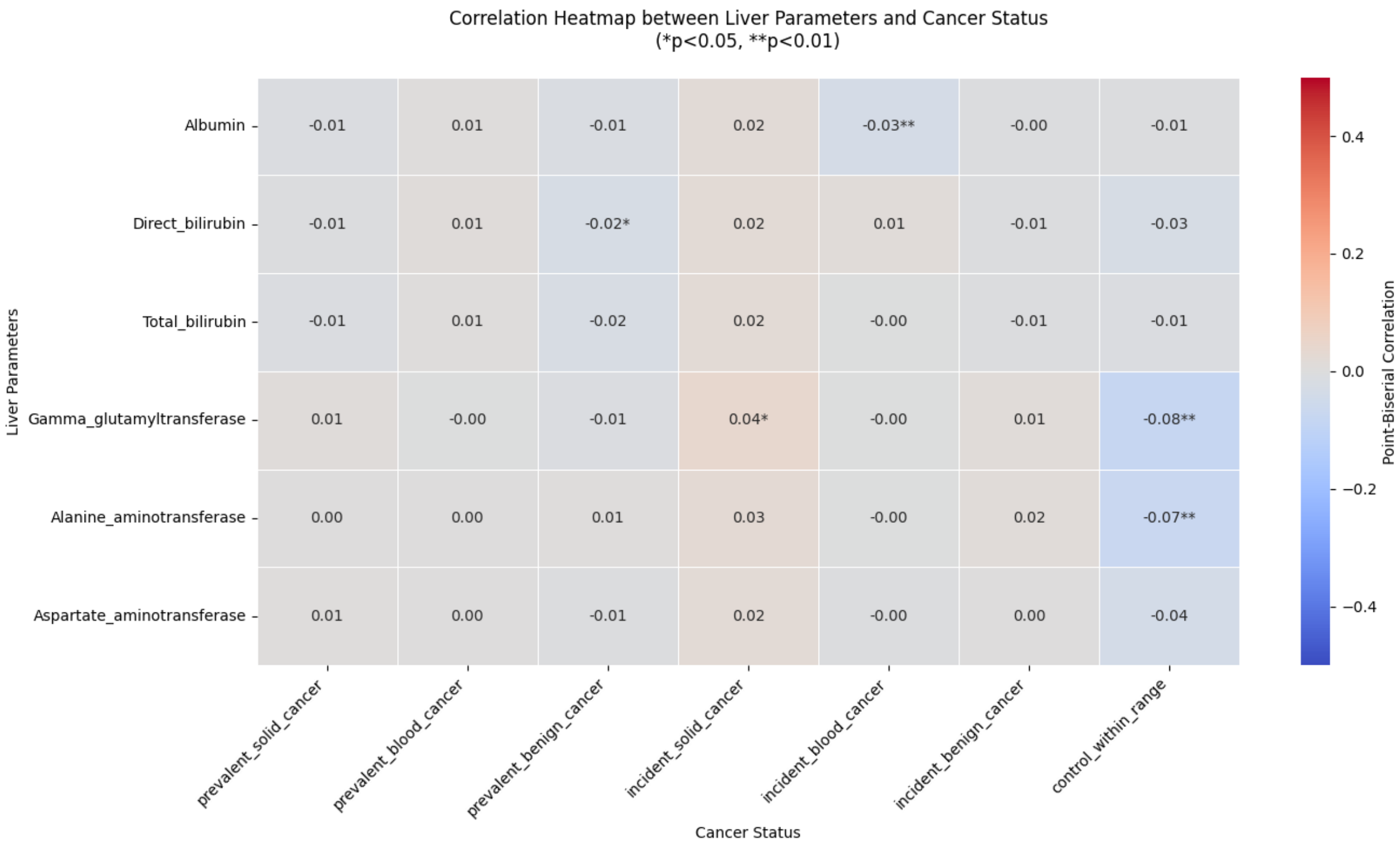

**Figure 3.10** Correlation Heatmap between Liver Parameters and Cancer Status

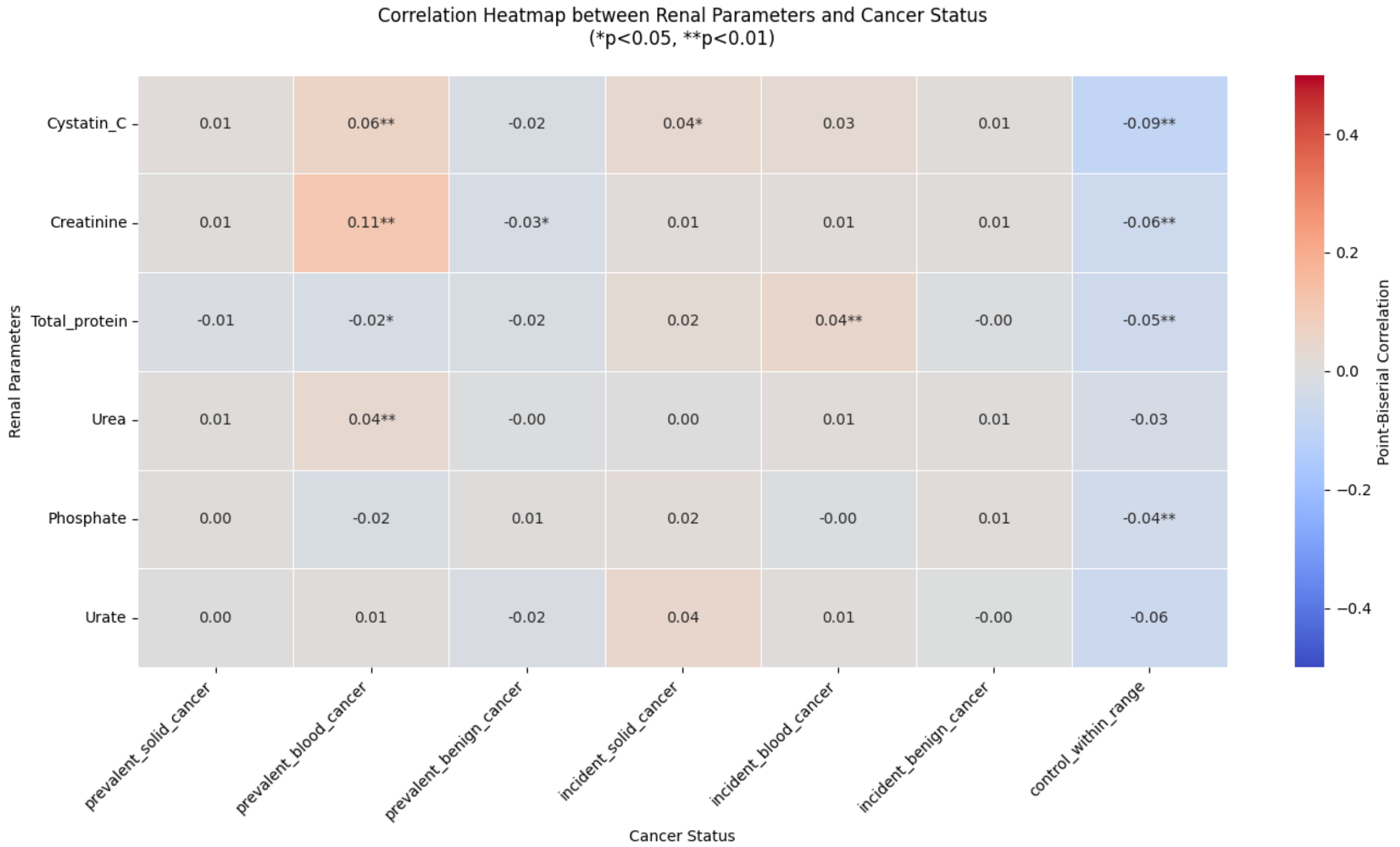

**Figure 3.11** Correlation Heatmap between Renal Parameters and Cancer Status

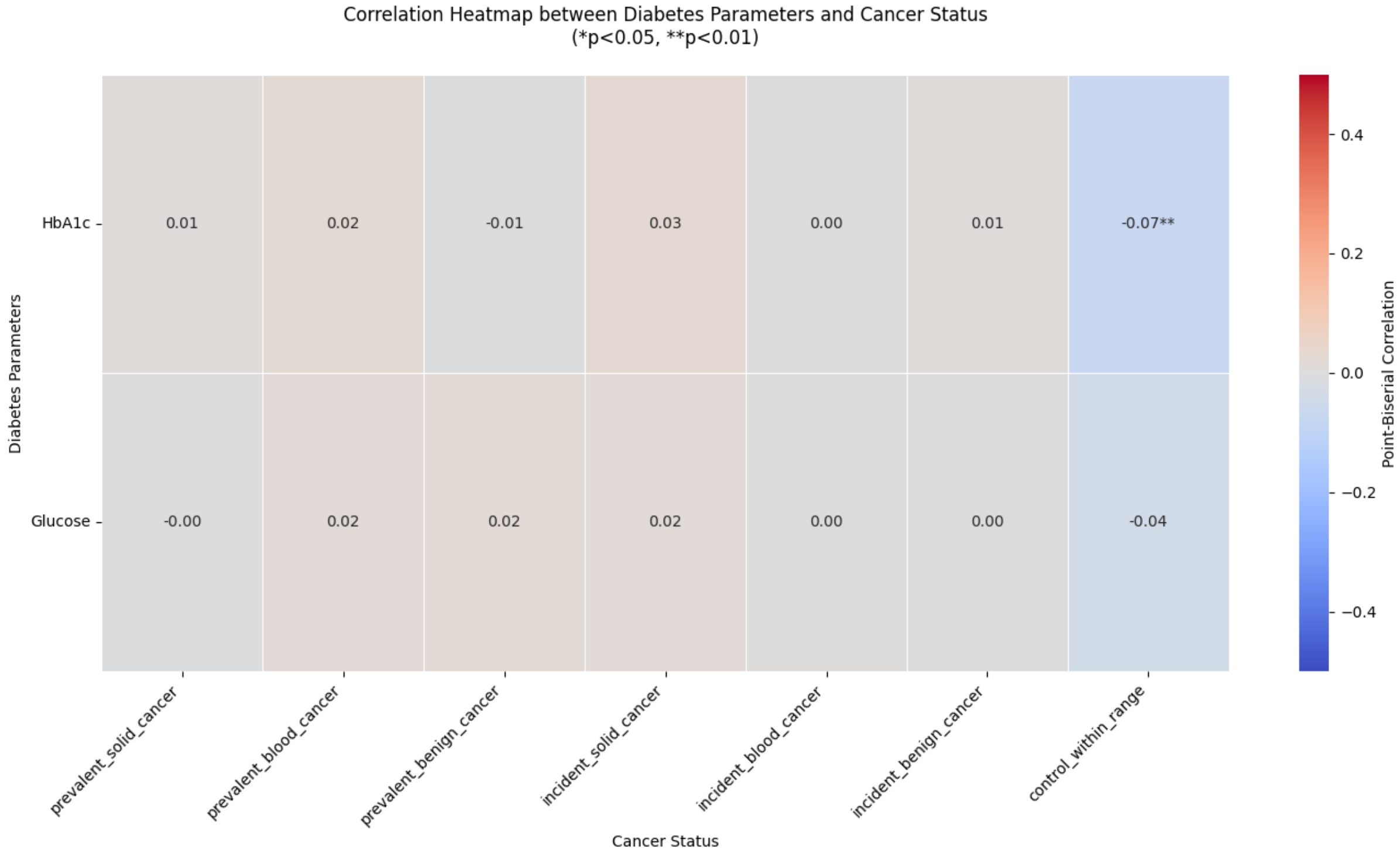

**Figure 3.12** Correlation Heatmap between Diabetes Parameters and Cancer Status

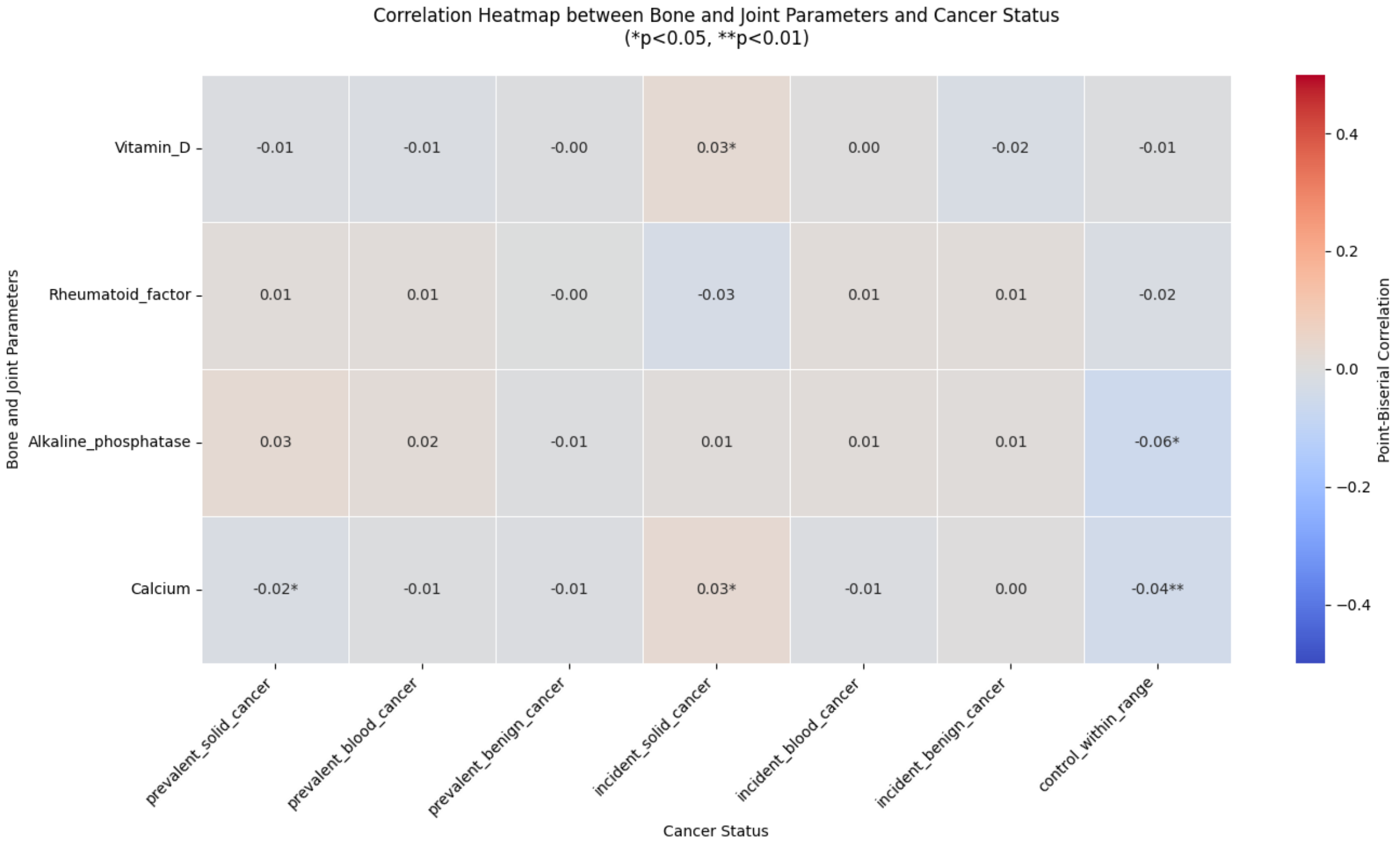

**Figure 3.13** Correlation Heatmap between Bone and Joint Parameters and Cancer Status

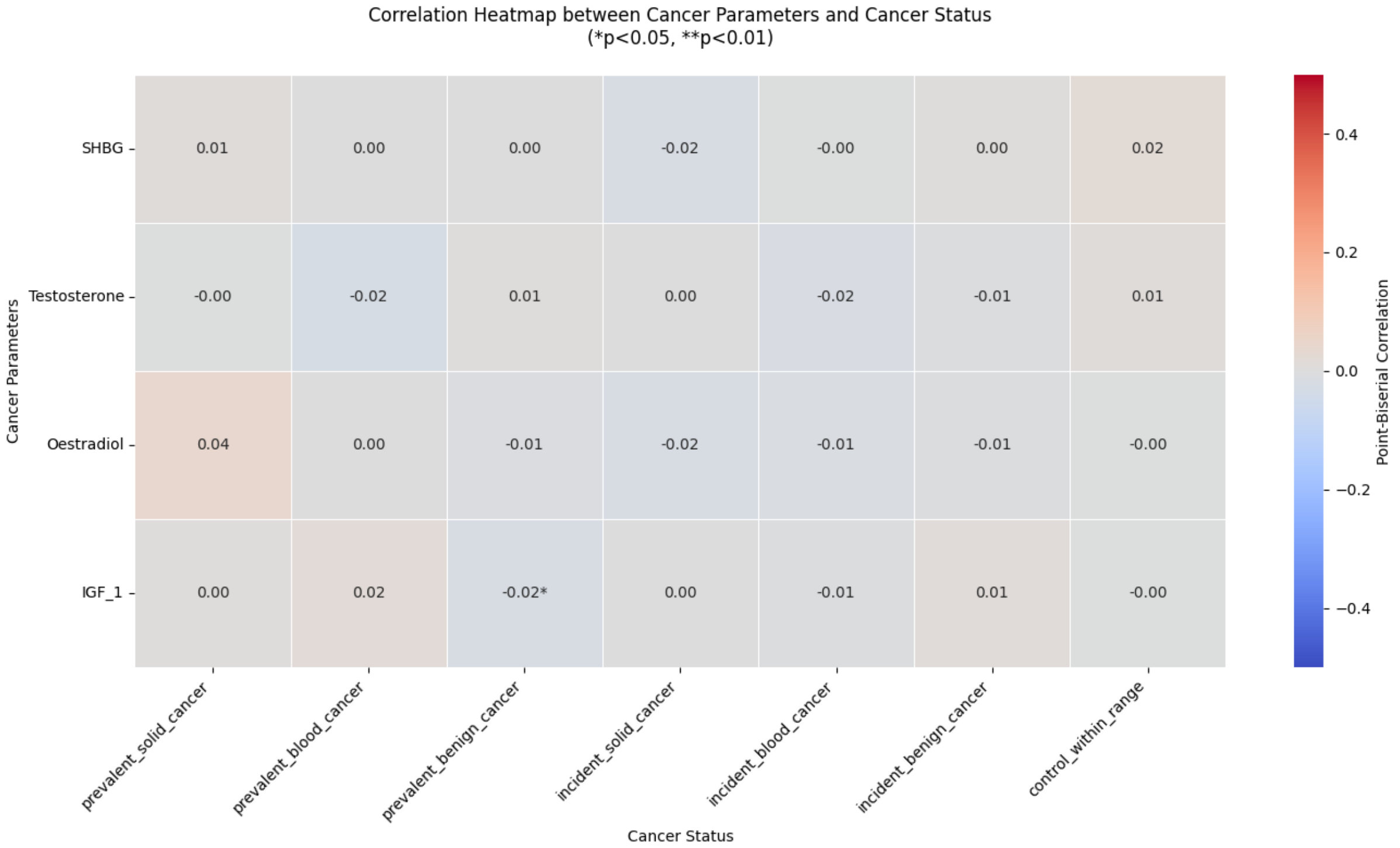

**Figure 3.14** Correlation Heatmap between Cancer Parameters and Cancer Status

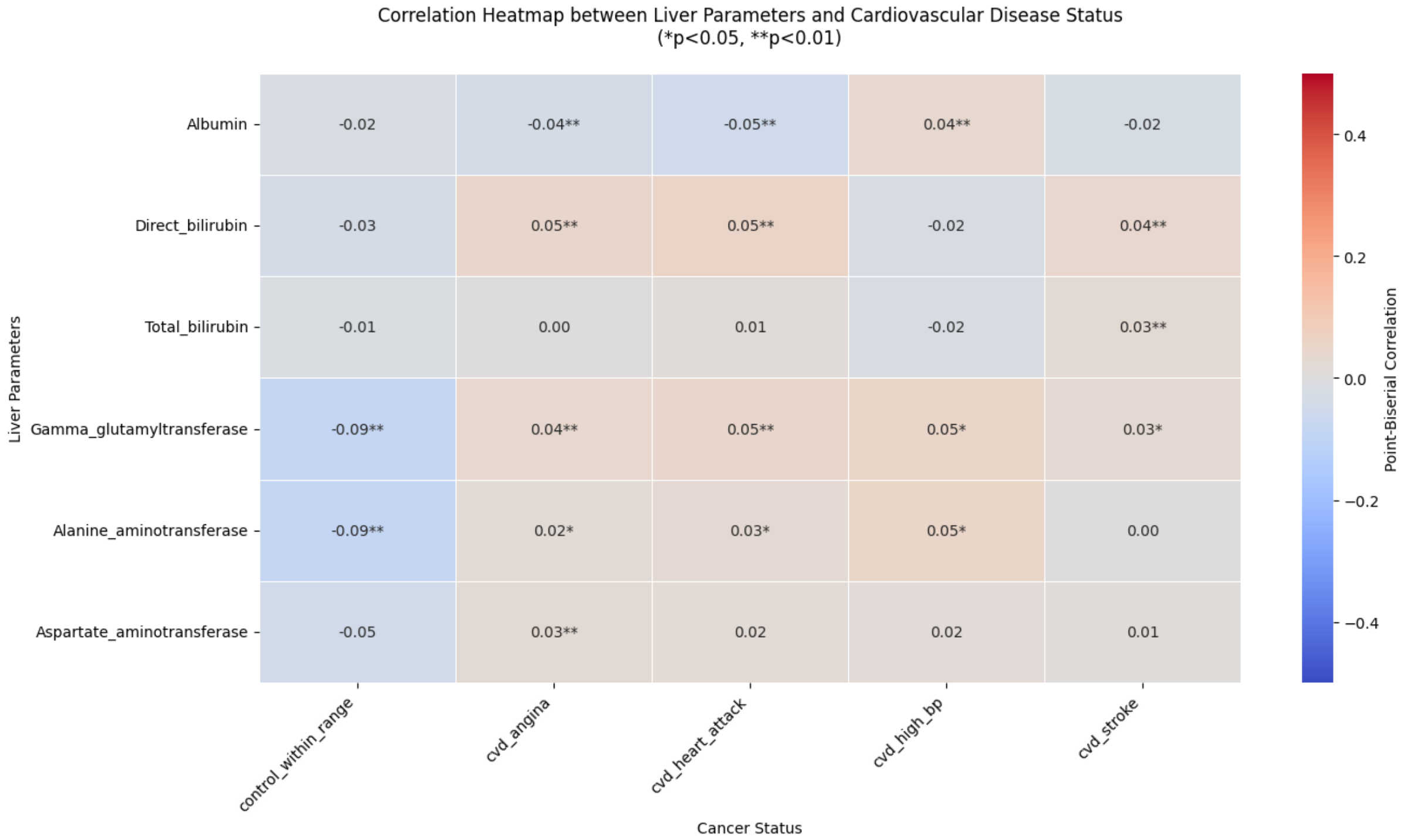

**Figure 3.17** Correlation Heatmap between Liver Parameters and Cancer Status

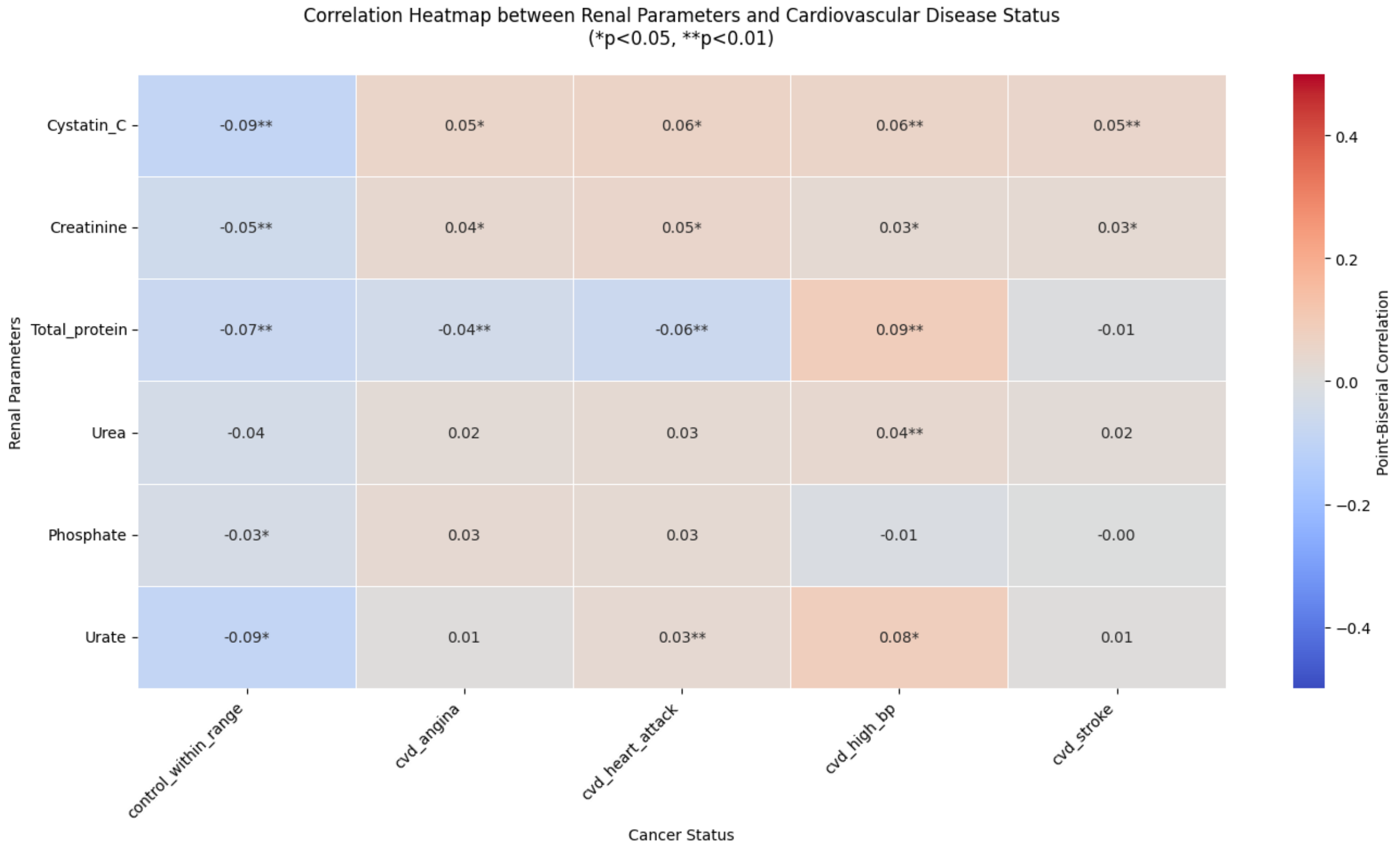

**Figure 3.18** Correlation Heatmap between Renal Parameters and Cancer Status

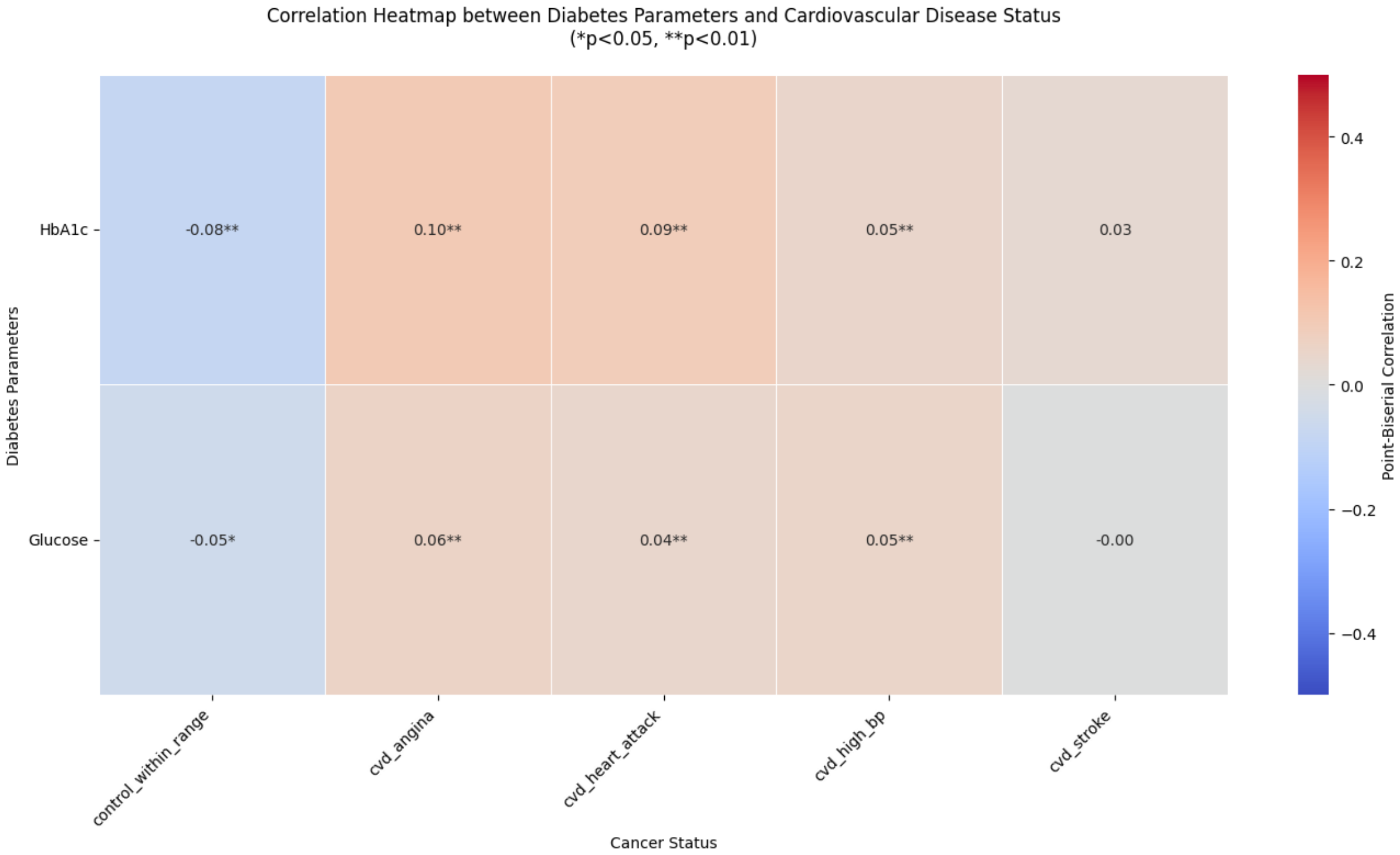

**Figure 3.19** Correlation Heatmap between Diabetes Parameters and Cancer Status

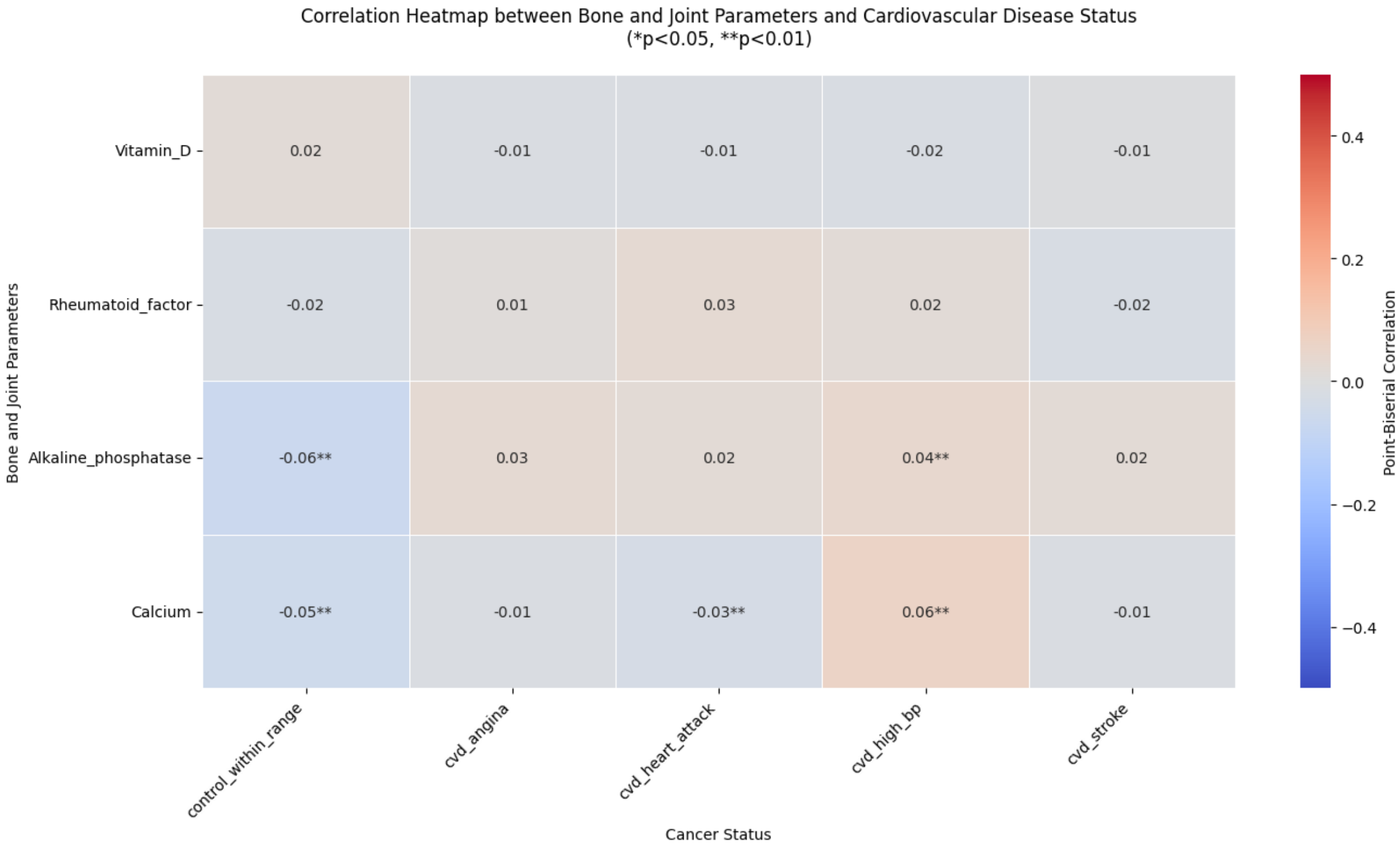

**Figure 3.20** Correlation Heatmap between Bone and Joint Parameters and Cancer Status

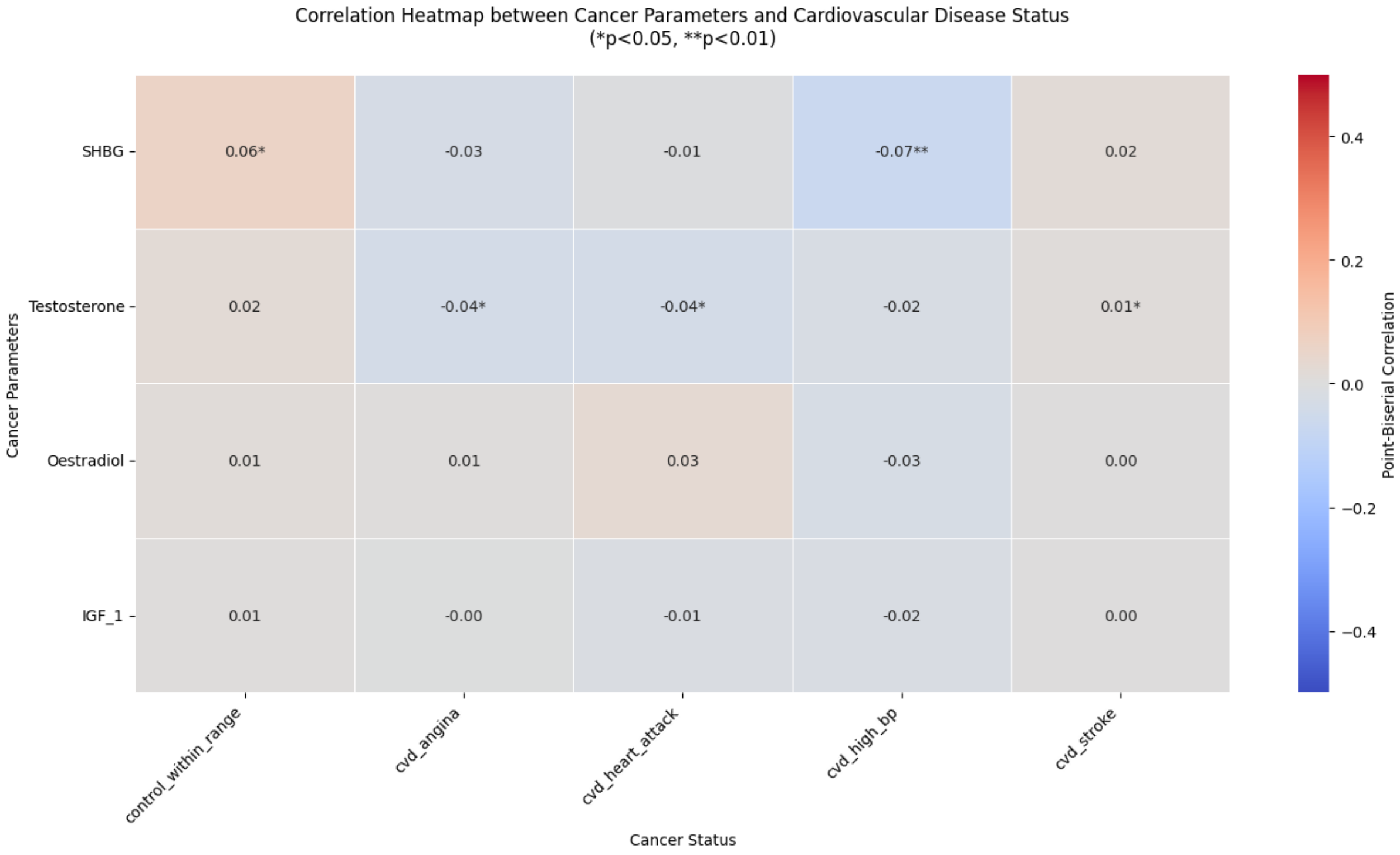

**Figure 3.21** Correlation Heatmap between Cancer Parameters and Cancer Status

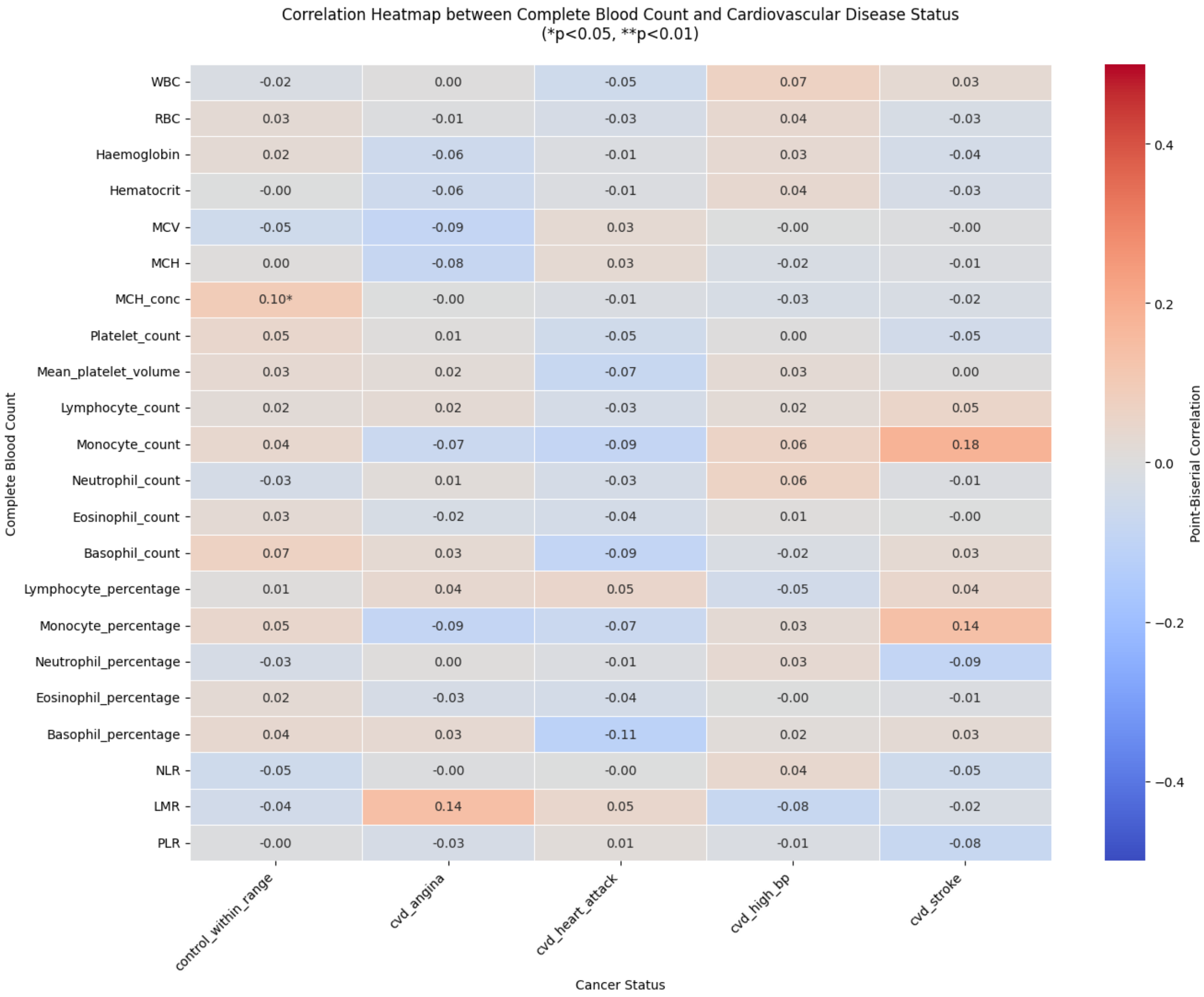

**Figure 3.22** Correlation Heatmap between CBC and Cancer Status

## MODEL PERFORMANCE TABLES

| Cancer Type | ROC-AUC | PR-AUC | Threshold | Precision | Recall | Specificity | F1 | Positives (train) |
|---|---|---|---|---|---|---|---|---|
| Prevalent Solid Cancer | 0.6287 | 0.3587 | 0.370 | 0.3094 | 0.8074 | 0.3652 | 0.4474 | 20129 |
| Prevalent Blood Cancer | 0.5695 | 0.0230 | 0.339 | 0.0582 | 0.0466 | 0.9908 | 0.0518 | 964 |
| Prevalent Benign Cancer | 0.6372 | 0.0945 | 0.420 | 0.0989 | 0.3573 | 0.8135 | 0.1549 | 4339 |
| Incident Solid Cancer | 0.6774 | 0.6336 | 0.314 | 0.5643 | 0.9316 | 0.2834 | 0.7029 | 38782 |
| Incident Blood Cancer | 0.5680 | 0.0581 | 0.459 | 0.0877 | 0.1084 | 0.9551 | 0.0969 | 2809 |
| Incident Benign Cancer | 0.5420 | 0.0881 | 0.347 | 0.0886 | 0.5268 | 0.5462 | 0.1517 | 6508 |

**Table S1.** Per-label XGBoost model (model 1) performance before hyperparameter tuning

| Cancer Type | ROC-AUC | PR-AUC | Threshold | Precision | Recall | Specificity | F1 |
|---|---|---|---|---|---|---|---|
| Prevalent Solid Cancer | 0.6343 | 0.3637 | 0.223 | 0.3161 | 0.7653 | 0.4167 | 0.4474 |
| Prevalent Blood Cancer | 0.6970 | 0.0521 | 0.066 | 0.1458 | 0.1186 | 0.9915 | 0.1308 |
| Prevalent Benign Cancer | 0.6911 | 0.1089 | 0.082 | 0.1129 | 0.4159 | 0.8127 | 0.1776 |
| Incident Solid Cancer | 0.6770 | 0.6294 | 0.333 | 0.5809 | 0.8991 | 0.3536 | 0.7058 |
| Incident Blood Cancer | 0.6262 | 0.0781 | 0.062 | 0.1018 | 0.1700 | 0.9403 | 0.1273 |

| Cancer Type | ROC-AUC | PR-AUC | Threshold | Precision | Recall | Specificity | F1 |
|---|---|---|---|---|---|---|---|
| Incident Benign Cancer | 0.5717 | 0.0950 | 0.086 | 0.0964 | 0.5341 | 0.5806 | 0.1633 |

**Table S2.** Per-label XGBoost model (model 1) performance after hyperparameter tuning

| Cancer Type | ROC-AUC | PR-AUC | Threshold | Precision | Recall | Specificity | F1 |
|---|---|---|---|---|---|---|---|
| Prevalent Solid Cancer | 0.6353 | 0.3993 | 0.462 | 0.3307 | 0.7303 | 0.4490 | 0.4552 |
| Prevalent Blood Cancer | 0.6941 | 0.0710 | 0.618 | 0.1128 | 0.1602 | 0.9839 | 0.1324 |
| Prevalent Benign Cancer | 0.7072 | 0.1362 | 0.621 | 0.1495 | 0.3396 | 0.8804 | 0.2076 |
| Incident Solid Cancer | 0.6655 | 0.6585 | 0.296 | 0.5834 | 0.9312 | 0.2518 | 0.7174 |
| Incident Blood Cancer | 0.7035 | 0.1506 | 0.529 | 0.1988 | 0.3324 | 0.9457 | 0.2488 |
| Incident Benign Cancer | 0.6003 | 0.1330 | 0.506 | 0.1273 | 0.4452 | 0.7006 | 0.1979 |

**Table S3.** Per-label XGBoost model (model 2) performance after hyperparameter tuning

## 3.2.5.2 Dimensionality Reduction Analysis

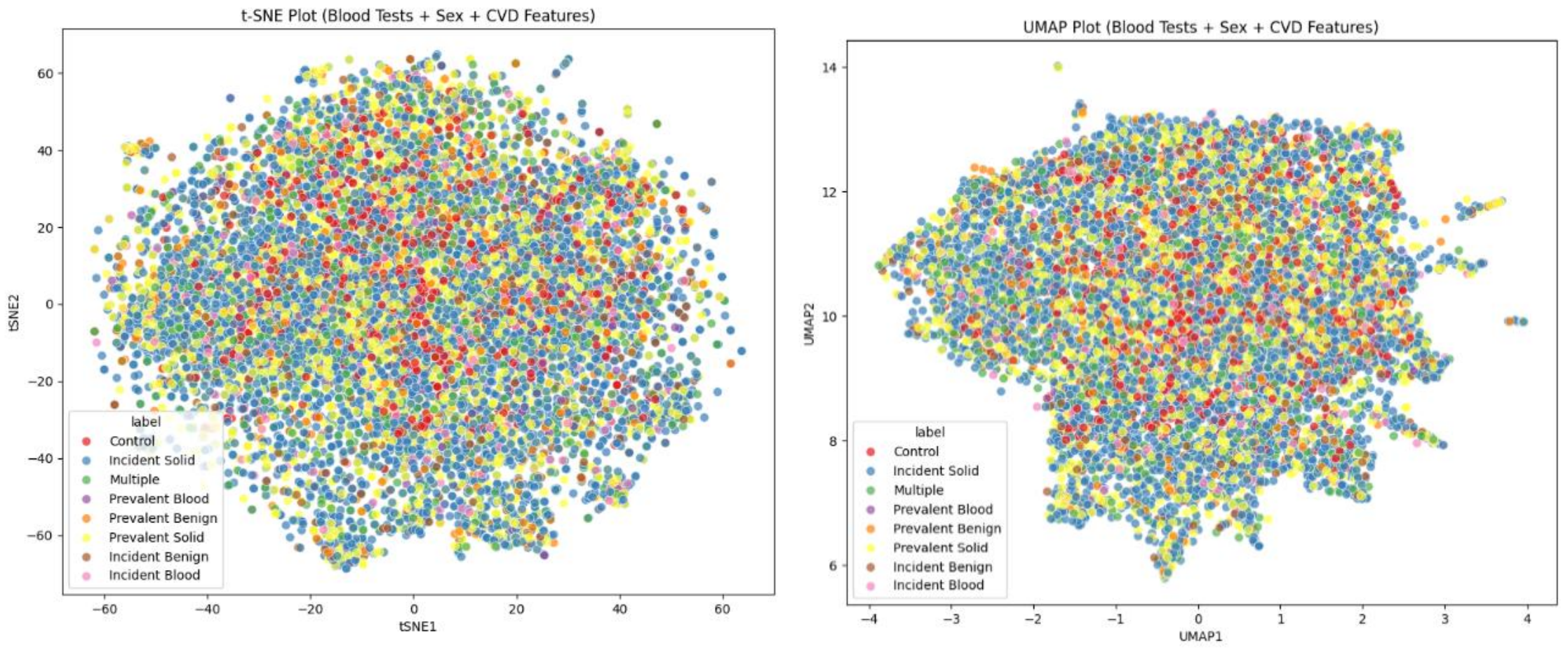

**Figure 3.24** Preliminary dimensionality reduction analysis results: (a) t-SNE (b) UMAP

For preliminary dimensionality reduction analysis, blood tests, sex and cardiovascular disease features were plotted against various cancer subtypes. The t-SNE and UMAP showed a single dense, roughly circular cluster with some outliers. There wasn't clear separation between different cancer groups and control groups, nor were there apparent clusters. The spread of points indicates that blood tests, sex and cardiovascular disease comorbidities may not be sufficient to distinguish various cancer statuses. More features such as age, prescribed medications and comorbidities could be incorporated to further improve the cancer prediction models.

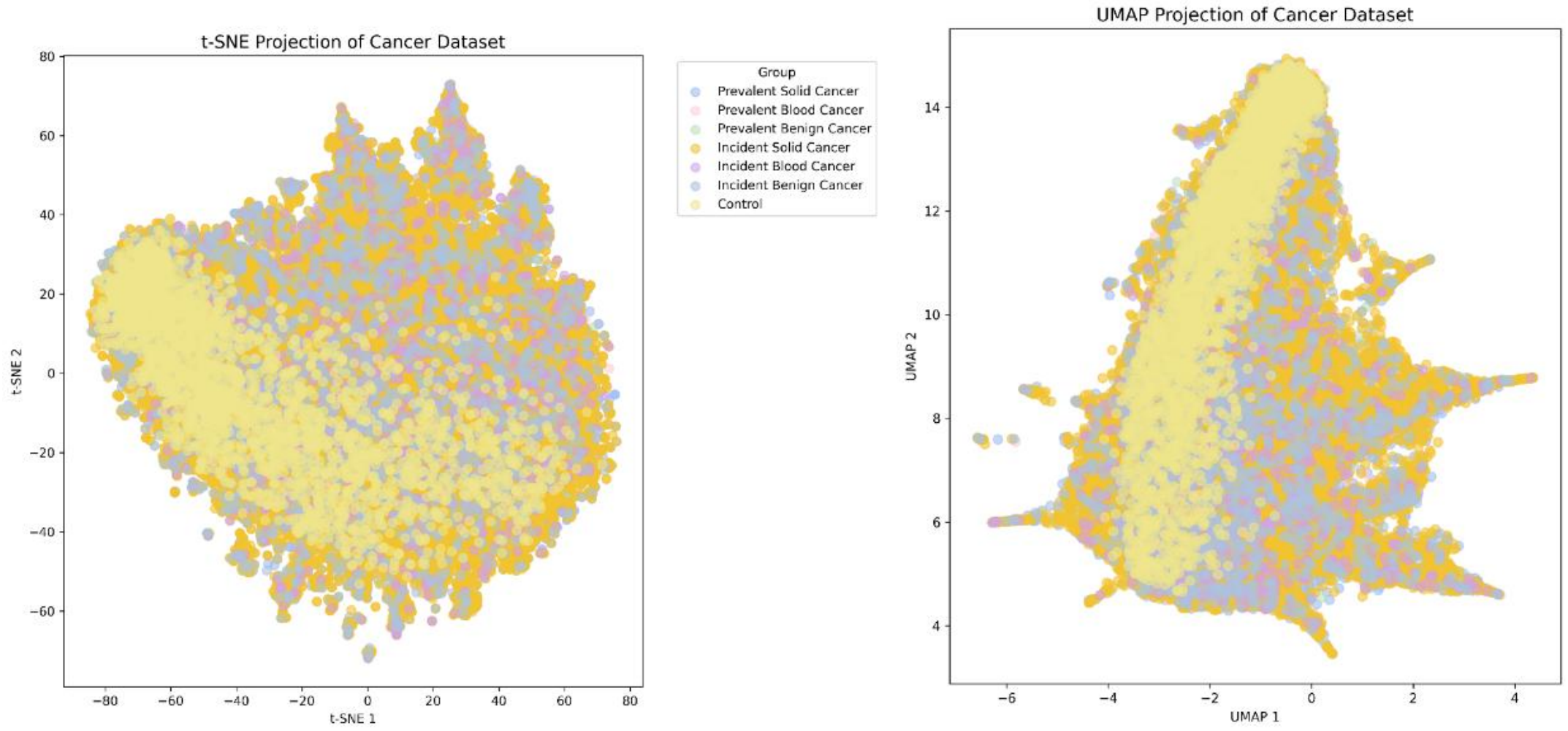

**Figure 3.25** Final dimensionality reduction analysis results: (a) t-SNE (b) UMAP

Dimensionality reduction analysis was performed using the final dataset for training th final cancer prediction model. Features incorporated include z-score of blood test parameters, age, sex, prescribed medications and comorbidities. There was a clear cluster of control features in t-SNE and UMAP. However, there were no clear clusters of cancer cases. There was significant overlapping between incident solid cancer, prevalent solid cancer and incident blood cancer features. This could explain the poor ability of the cancer prediction models in differentiating between different types of cancer cases and control.

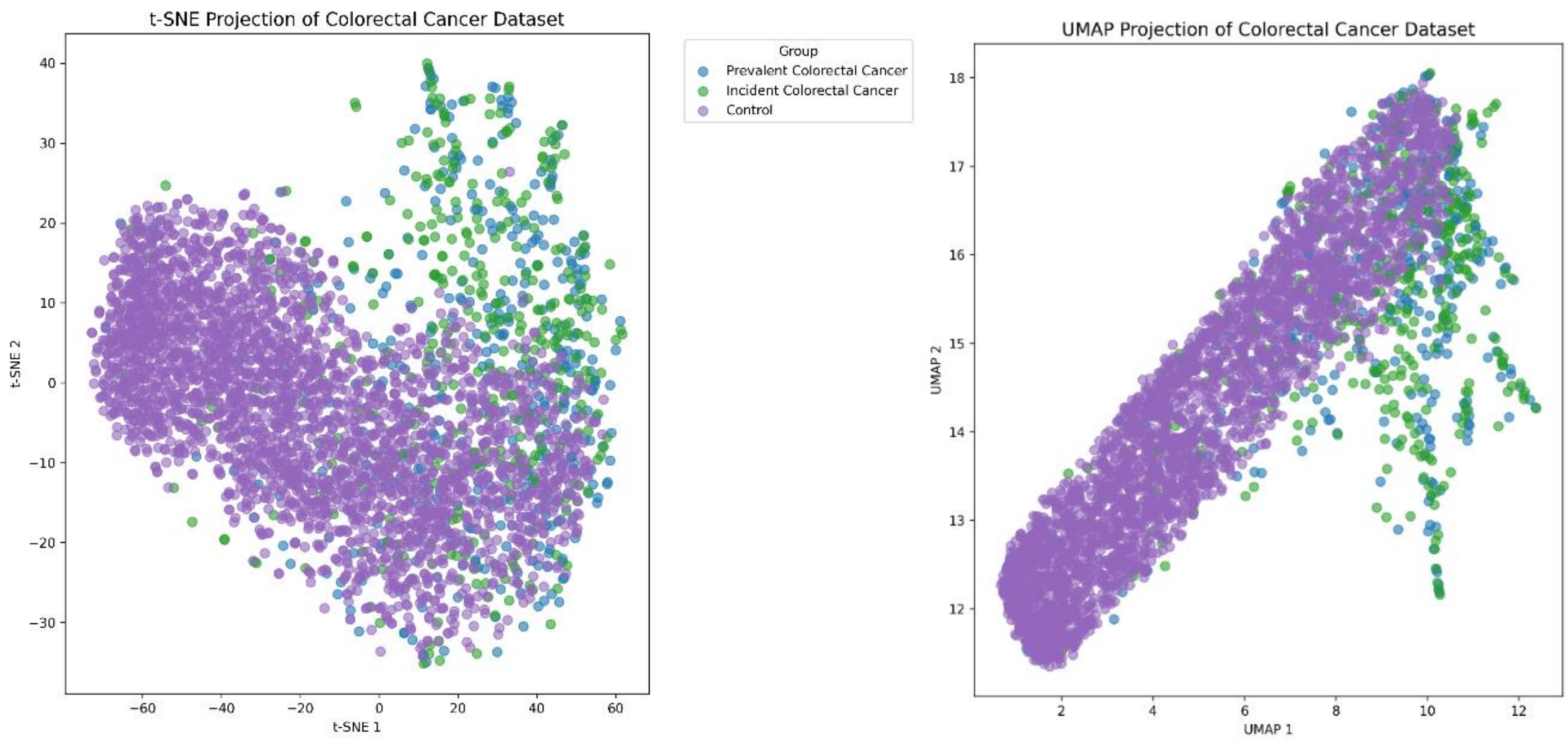

**Figure 3.28** Dimensionality reduction analysis results: (a) t-SNE (b) UMAP

The t-SNE and UMAP showed clear clustering of the control group and colorectal cancer subjects. The control group subjects formed a dense and regular cluster (purple), demonstrating homogeneity in healthy individuals' blood test, comorbidities and prescribed medication profile. Most colorectal cancer subjects deviated from the control group cluster, but there was still a minority of colorectal cancer subjects co-localizing with control group subjects. There was significant overlap in the projection of incident and prevalent colorectal cancer subjects, which may pose difficulty in differentiating between prevalent and incident cancers.

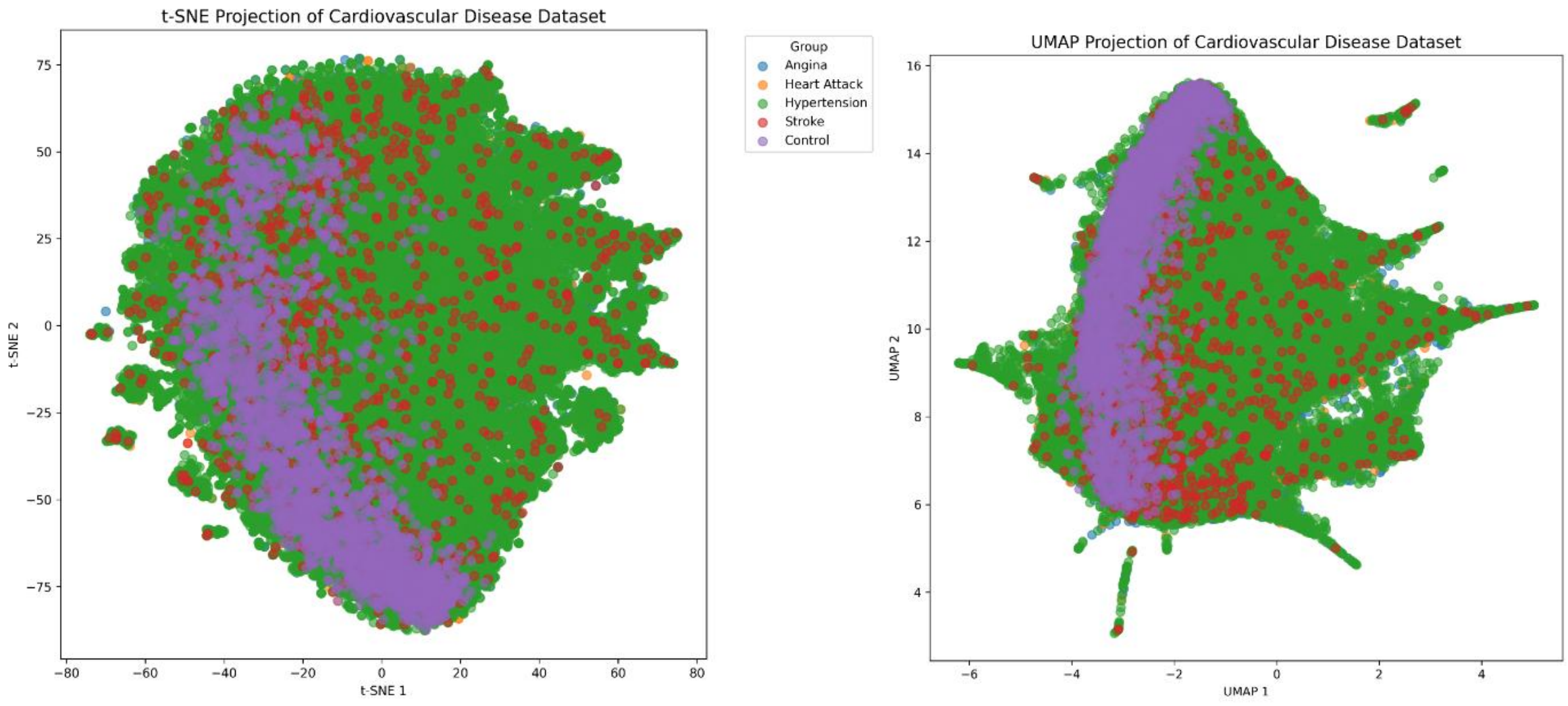

**Figure 3.31** Dimensionality reduction analysis results: (a) t-SNE (b) UMAP

Reviewing the t-SNE and UMAP, there was a clear cluster of control group features. However, there were no clear clusters for each of the cardiovascular diseases. Given the vast number of hypertension cases, the features of hypertension cases were dispersed around the projection. There was significant overlap between stroke and hypertension feature projections. Angina and heart attack features were scattered at the edges of the projection. Hence, the accuracy of models to predict angina, heart attack and stroke was not as high as predicting hypertension, which had more cases.

| Cancer type | Prevalent | Incident |
|---|---|---|
| Solid | 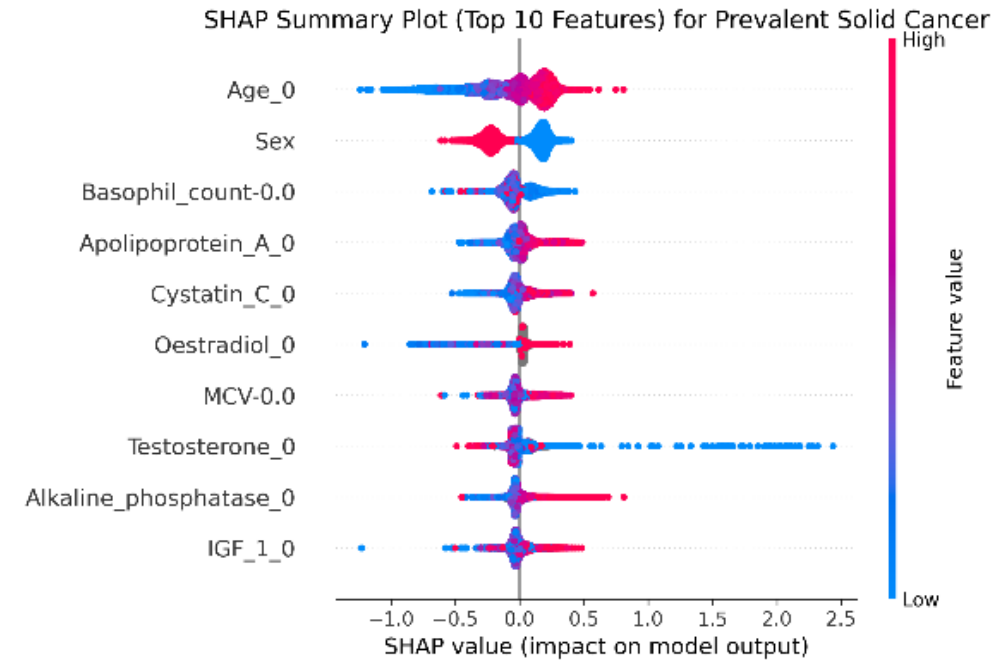 | 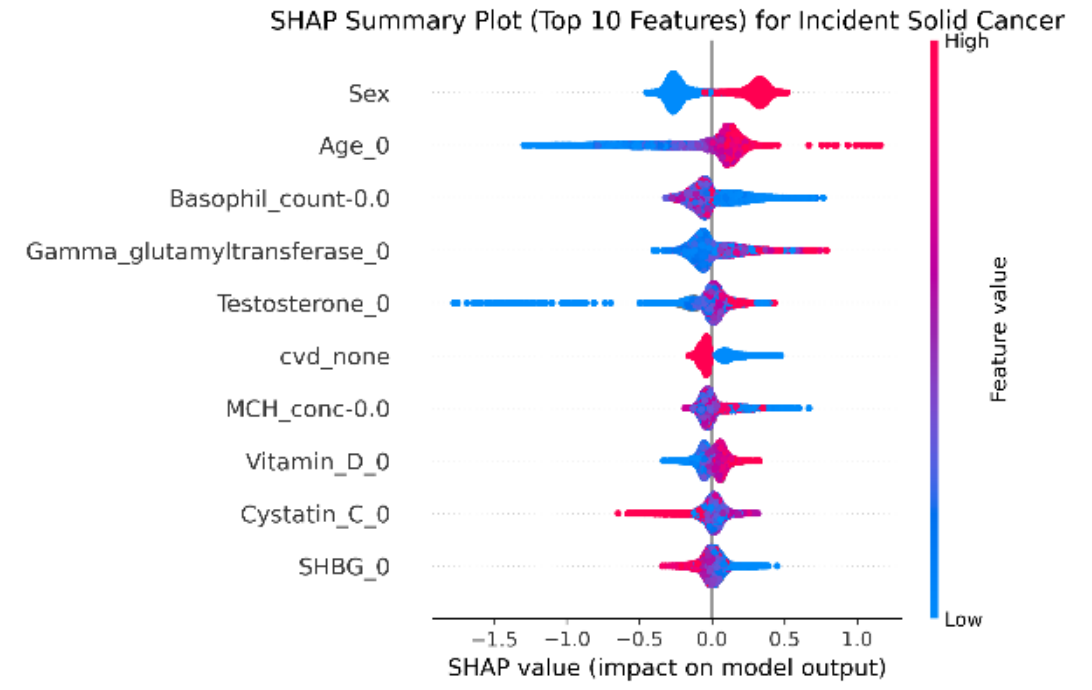 |
| Blood | 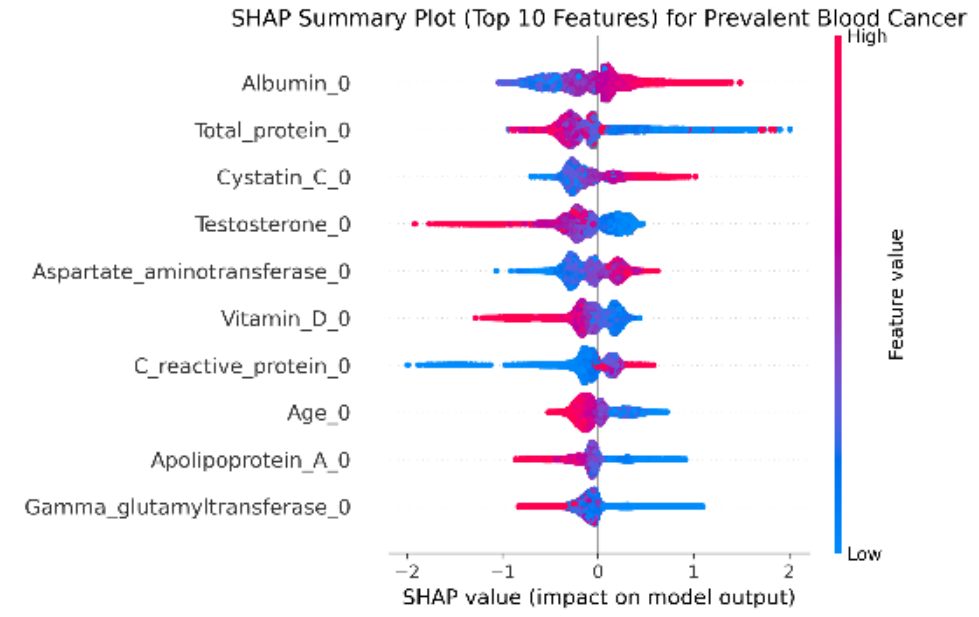 | 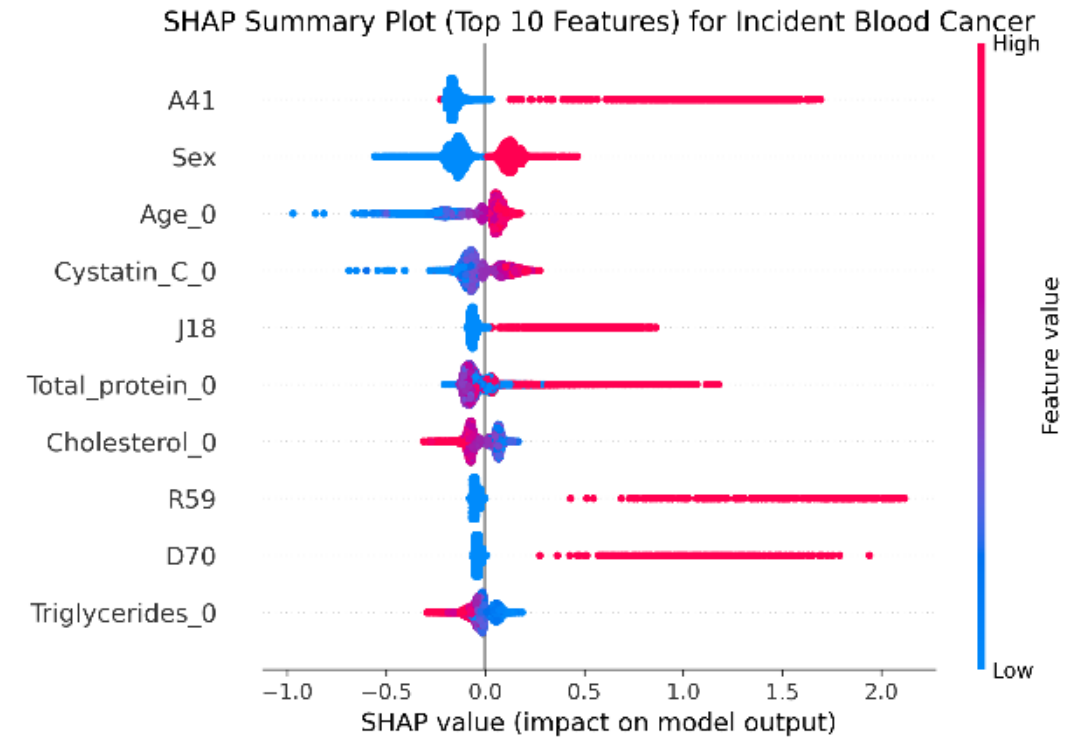 |
| Benign | 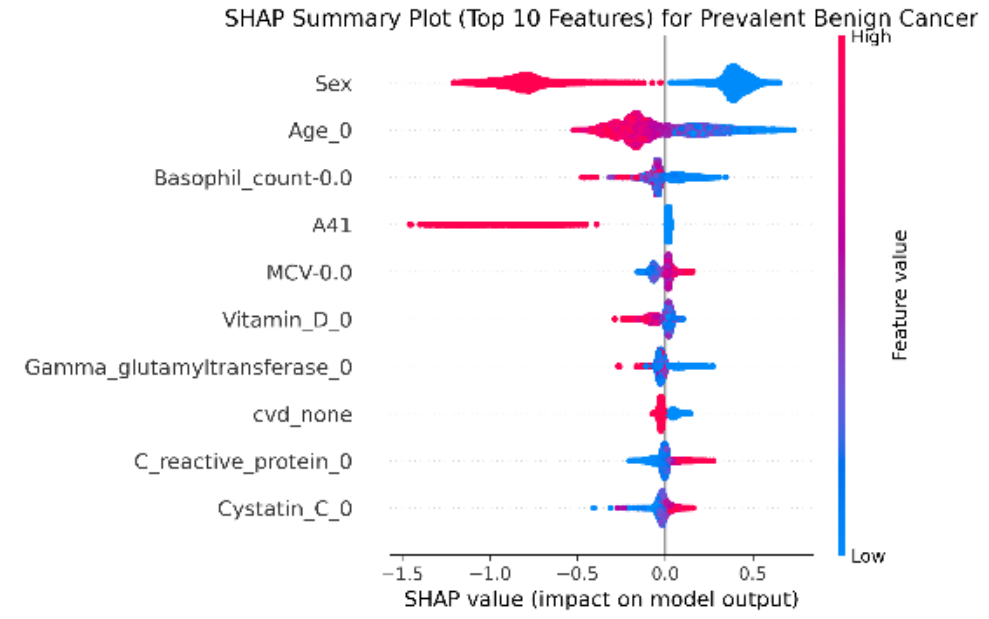 | 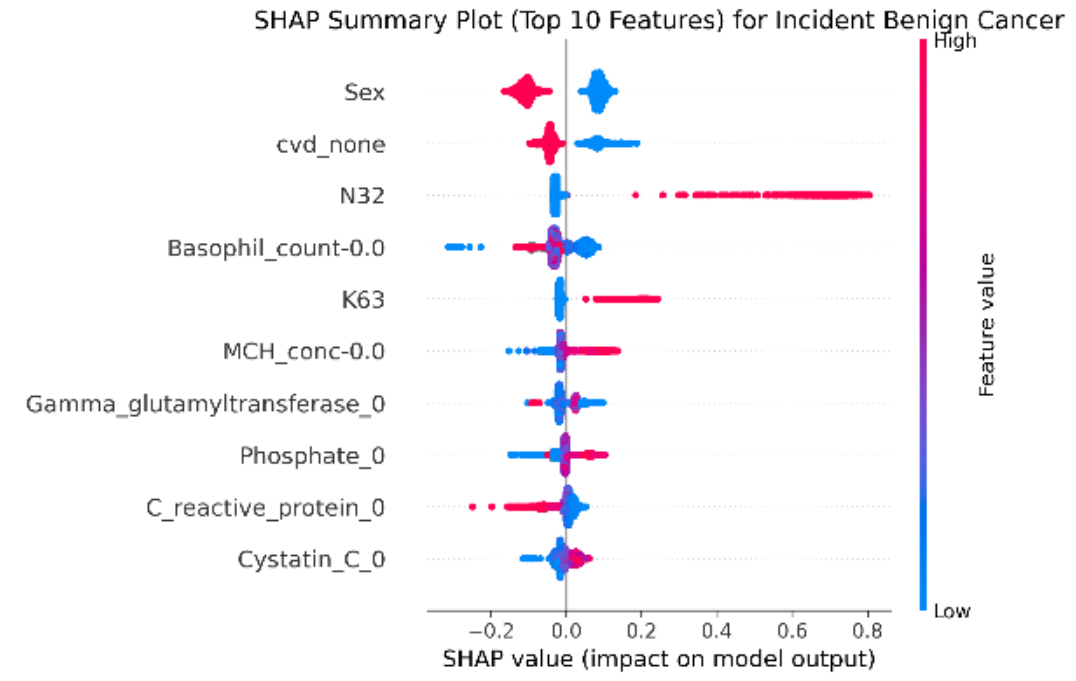 |

**Table 3.9** SHAP Summary Plots for cancer prediction

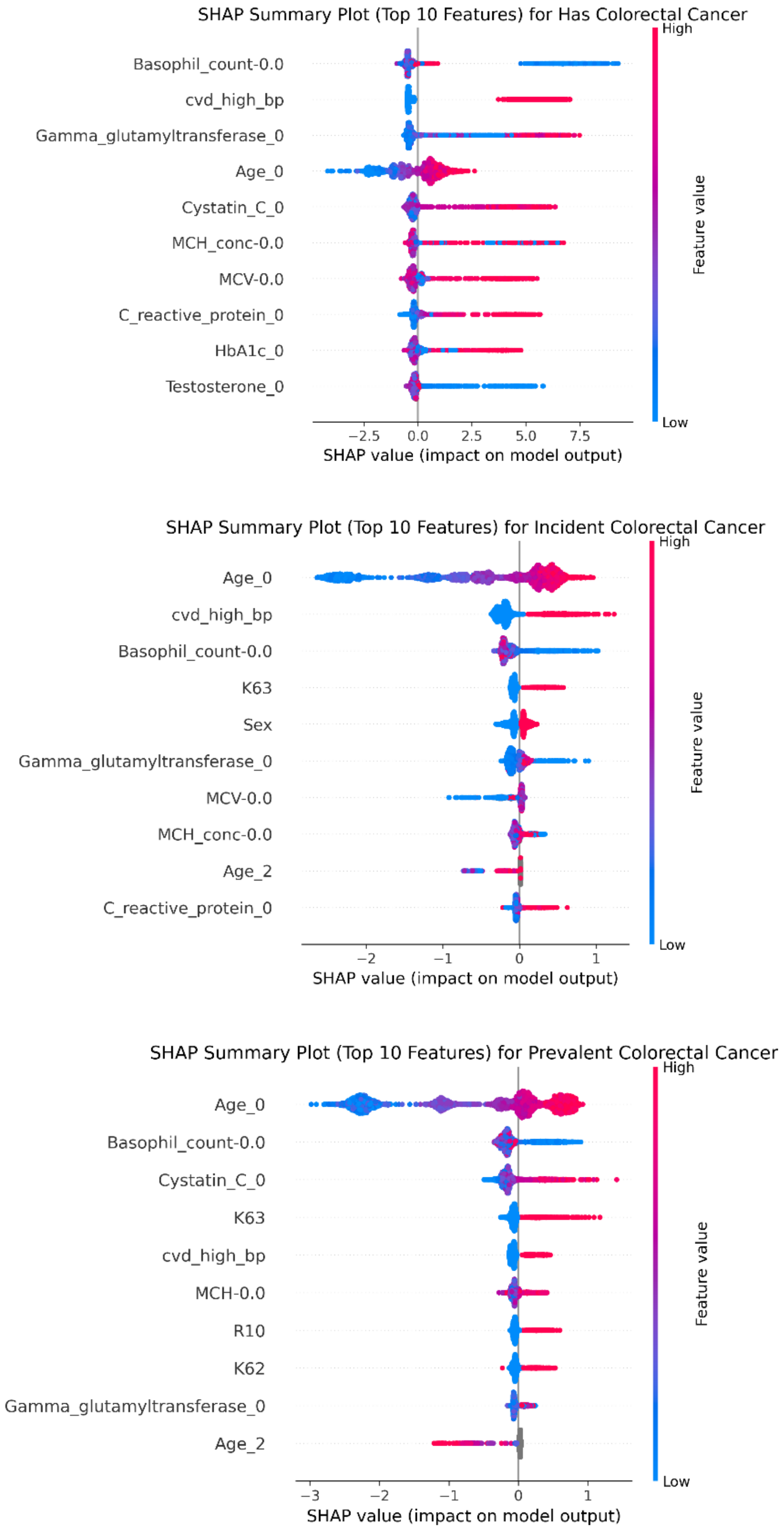

**Figure 3.27** SHAP summary plots for colorectal cancer prediction

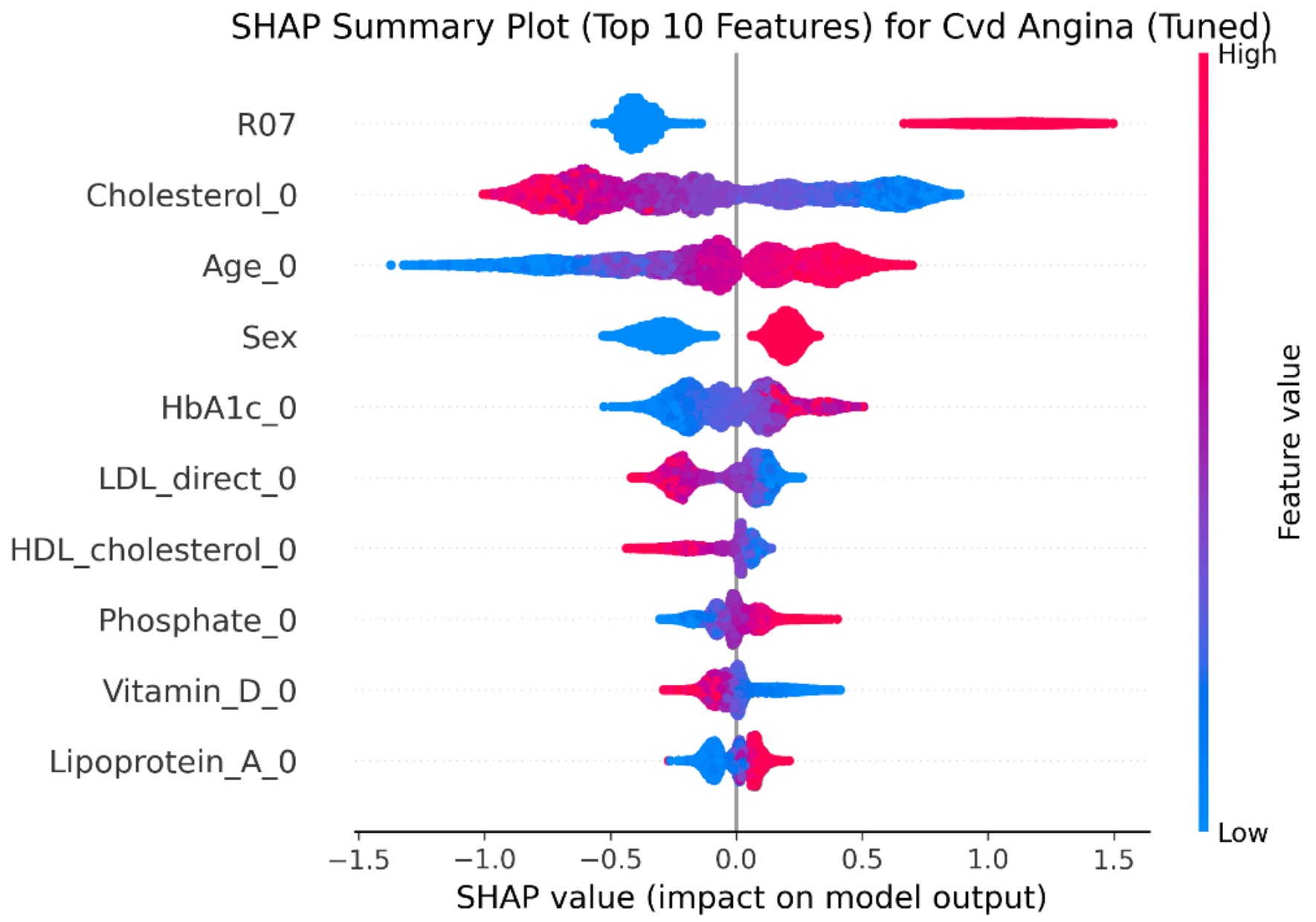
SHAP Summary Plot (Top 10 Features) for Cvd Angina (Tuned)
R07
Cholesterol_0
Age_0
Sex
HbA1c_0
LDL_direct_0
HDL_cholesterol_0
Phosphate_0
Vitamin_D_0
Lipoprotein_A_0
−1.5 −1.0 −0.5 0.0 0.5 1.0 1.5
SHAP value (impact on model output)
High
Feature value
Low

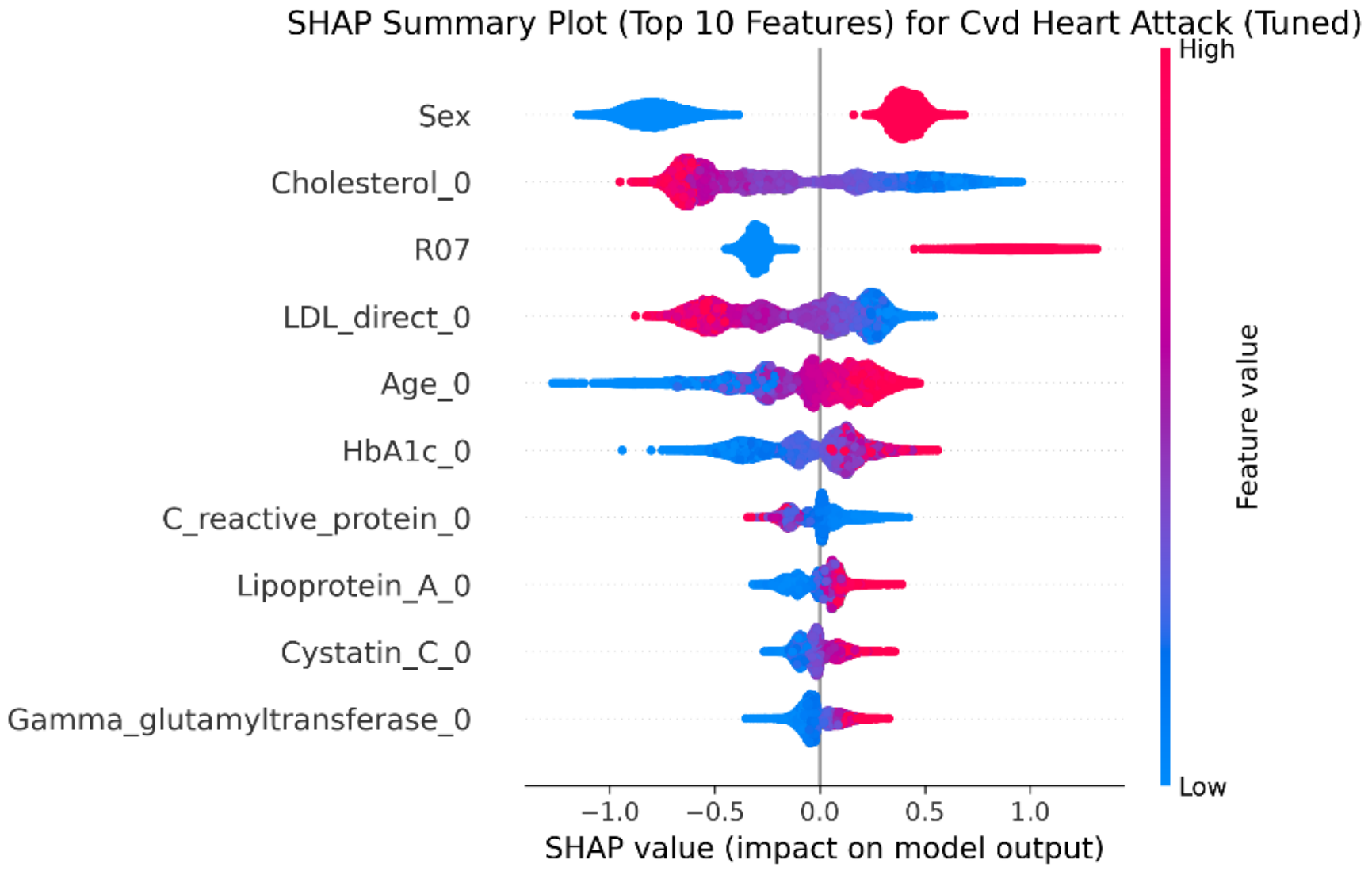
SHAP Summary Plot (Top 10 Features) for Cvd Heart Attack (Tuned)
Sex
Cholesterol_0
R07
LDL_direct_0
Age_0
HbA1c_0
C_reactive_protein_0
Lipoprotein_A_0
Cystatin_C_0
Gamma_glutamyltransferase_0
−1.0 −0.5 0.0 0.5 1.0
SHAP value (impact on model output)
High
Feature value
Low

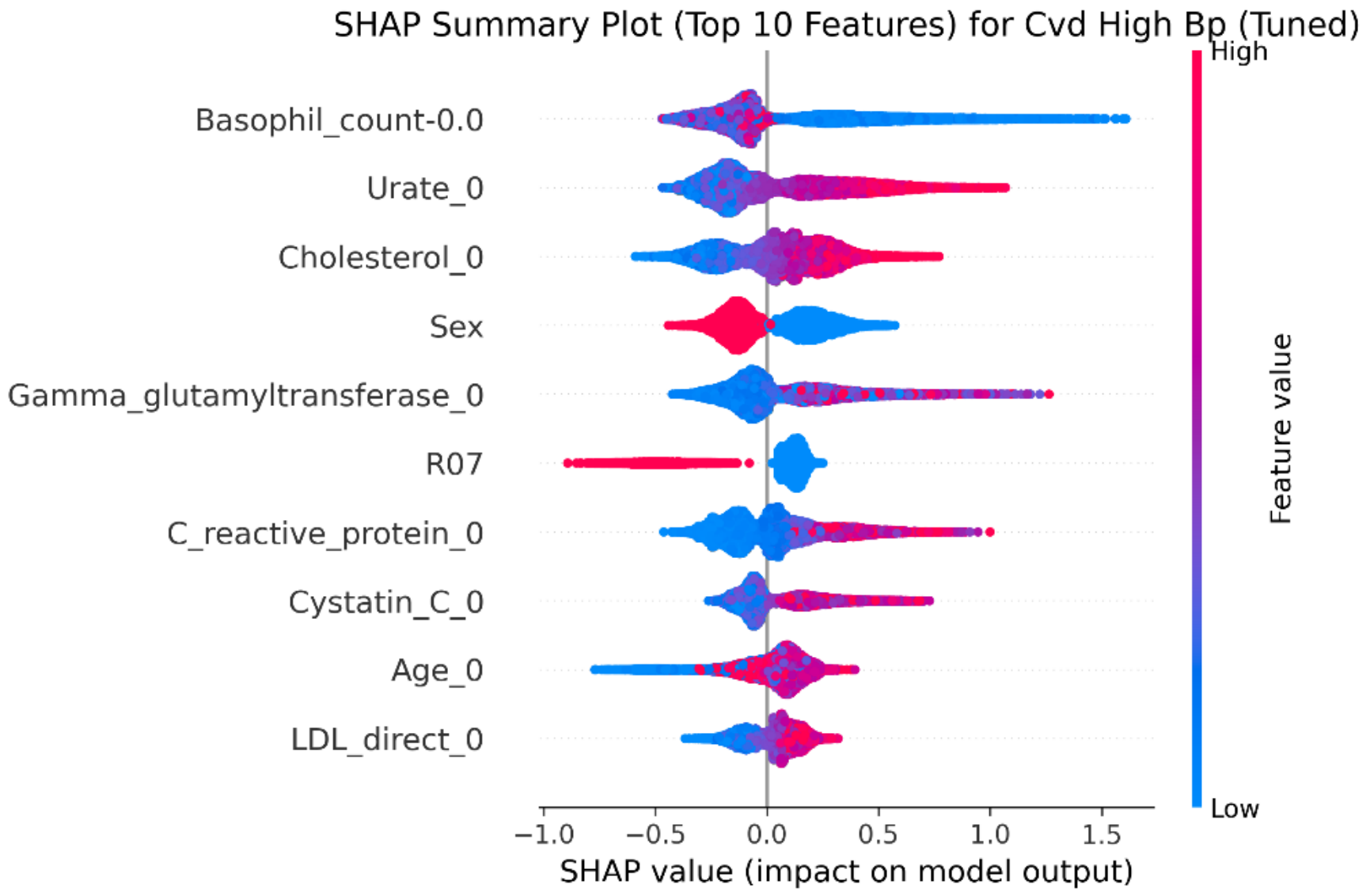
SHAP Summary Plot (Top 10 Features) for Cvd High Bp (Tuned)
Basophil_count-0.0
Urate_0
Cholesterol_0
Sex
Gamma_glutamyltransferase_0
R07
C_reactive_protein_0
Cystatin_C_0
Age_0
LDL_direct_0
−1.0 −0.5 0.0 0.5 1.0 1.5
SHAP value (impact on model output)
High
Feature value
Low

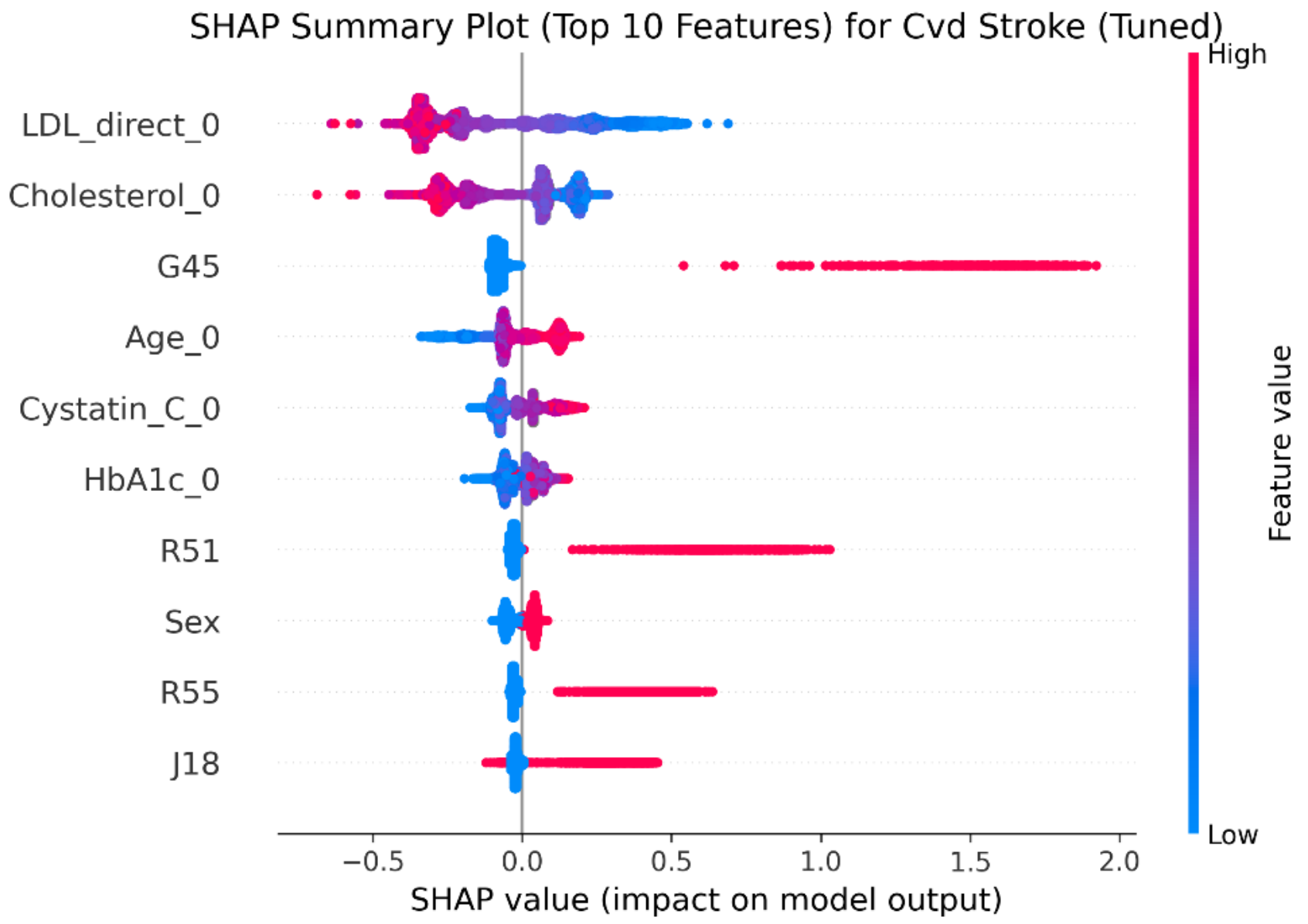

**Figure 3.30** SHAP summary plots for cardiovascular disease prediction

| Parameter | Unit | Effect Size | P value | Mean (Pre-flight) | Mean (Post-flight) | p_adj |
|---|---|---|---|---|---|---|
| White blood cell count | Thousand/uL | -0.65556 | 0.076871 | 6.855556 | 6.2 | 0.384353 |
| Red blood cell count | Million/uL | 0.040556 | 0.248878 | 4.731111 | 4.771667 | 0.441826 |
| Hemoglobin | g/dL | -0.16389 | 0.396038 | 14.62222 | 14.45833 | 0.609289 |
| Hematocrit | % | -1.01389 | 0.104884 | 43.85556 | 42.84167 | 0.411135 |
| MCV | fL | -3.18056 | 0.035651 | 93.01111 | 89.83056 | 0.384353 |
| MCH | pg | -0.56667 | 0.123341 | 30.88333 | 30.31667 | 0.411135 |
| MCHC | g/dL | 0.4375 | 0.171552 | 33.55 | 33.9875 | 0.428879 |
| RDW | % | -0.17778 | 0.436249 | 12.67778 | 12.5 | 0.623213 |
| Platelet count | Thousand/uL | -4.25 | 0.636766 | 245.3333 | 241.0833 | 0.749136 |
| MPV | fL | 0.088889 | 0.22886 | 10.34444 | 10.43333 | 0.441826 |
| Absolute neutrophils | cells/uL | -581.694 | 0.042764 | 3775.444 | 3193.75 | 0.384353 |
| Absolute lymphocytes | cells/uL | -104.306 | 0.511875 | 2231.889 | 2127.583 | 0.643225 |
| Absolute monocytes | cells/uL | 32.44444 | 0.51458 | 578.2222 | 610.6667 | 0.643225 |
| Absolute eosinophils | cells/uL | -2.19444 | 0.964441 | 212.7778 | 210.5833 | 0.965718 |
| Absolute basophils | cells/uL | 0.111111 | 0.965718 | 57.55556 | 57.66667 | 0.965718 |
| Neutrophils | % | -3.21667 | 0.164005 | 54.43333 | 51.21667 | 0.428879 |
| Lymphocytes | % | 1.791667 | 0.237718 | 33.2 | 34.99167 | 0.441826 |
| Monocytes | % | 1.216667 | 0.063577 | 8.3 | 9.516667 | 0.384353 |
| Eosinophils | % | 0.119444 | 0.858451 | 3.222222 | 3.341667 | 0.953834 |
| Basophils | % | 0.088889 | 0.265096 | 0.844444 | 0.933333 | 0.441826 |

**Table S4.** Paired t-test results for CBC before and after spaceflight

## Prescribed medications

| Medication | Number of patients prescribed | Percentage of patients prescribed | Number of prescriptions over 3 timepoints | Percentage of prescriptions over 3 timepoints |
|---|---|---|---|---|
| paracetamol | 98182 | 26.30321215 | 101890 | 7.373320578 |
| aspirin | 69410 | 18.59511881 | 72890 | 5.27472114 |
| ibuprofen | 65267 | 17.48519838 | 67301 | 4.870270372 |
| simvastatin | 60802 | 16.28901331 | 64380 | 4.658890753 |
| omeprazole | 33713 | 9.031800038 | 35836 | 2.593289982 |
| glucosamine product | 33402 | 8.948482332 | 34189 | 2.474104007 |
| cod liver oil capsule | 29314 | 7.853296541 | 29930 | 2.165899351 |
| bendroflumethiazide | 28713 | 7.692287084 | 30049 | 2.174510845 |
| ramipril | 25514 | 6.835266697 | 27393 | 1.982308083 |
| amlodipine | 24626 | 6.597369197 | 26317 | 1.904442807 |

**Table S5.** Top 10 most prescribed medications in the UK Biobank

As observed, the most commonly prescribed medication was paracetamol, with over a quarter of the patients having been prescribed for it at least once. Non-Steroidal Anti-Inflammatory Drugs (NSAIDs) such as aspirin (18.6%) and ibuprofen (17.5%) were the second- and third-most prescribed medications. The ten most prescribed medications and their usage frequencies are summarized (Table S5).

## *EDA on blood test parameters*

## Mean and standard deviation of blood test parameters

| Category | Blood Test Parameter | Cancer group | Cardiovascular Disease group | Final Control group | Unit |
|---|---|---|---|---|---|
| Demographics | Total number of subjects | 82826 | 112790 | 8845 | N/A |
| | Male | 38454 | 59720 | 3652 | N/A |
| | Female | 44372 | 53070 | 5193 | N/A |
| | Age | 60.28 ± 7.12 | 60.08 ± 7.08 | 53.84 ± 7.96 | Years |
| Cardiovascular Parameters | Cholesterol | 5.67 ± 1.18 | 5.38 ± 1.21 | 5.74 ± 0.89 | mmol/L |
| | Low-Density Lipoprotein direct (LDL direct) | 3.53 ± 0.89 | 3.33 ± 0.91 | 3.59 ± 0.69 | mmol/L |
| | High-Density Lipoprotein cholesterol (HDL cholesterol) | 1.45 ± 0.39 | 1.37 ± 0.37 | 1.50 ± 0.31 | mmol/L |
| | Triglycerides | 1.77 ± 1.00 | 1.92 ± 1.08 | 1.46 ± 0.77 | mmol/L |
| | Apolipoprotein A | 1.54 ± 0.28 | 1.50 ± 0.27 | 1.56 ± 0.22 | g/L |
| | Apolipoprotein B | 1.03 ± 0.24 | 0.99 ± 0.24 | 1.02 ± 0.19 | g/L |
| | C reactive protein | 2.85 ± 4.79 | 3.17 ± 4.93 | 1.23 ± 1.01 | mg/L |
| | Lipoprotein A | 44.57 ± 49.32 | 45.67 ± 50.08 | 44.89 ± 49.39 | nmol/L |
| Liver Parameters | Albumin | 44.98 ± 2.65 | 45.19 ± 2.71 | 45.39 ± 2.15 | g/L |
| | Direct bilirubin | 1.84 ± 0.85 | 1.90 ± 0.90 | 1.73 ± 0.58 | µmol/L |
| | Total bilirubin | 9.11 ± 4.40 | 9.26 ± 4.41 | 8.82 ± 3.17 | µmol/L |
| | Gamma glutamyltransferase (GGT) | 38.95 ± 46.17 | 44.94 ± 50.27 | 23.46 ± 6.46 | U/L |
| | Alanine aminotransferase (ALT) | 23.23 ± 13.67 | 25.89 ± 15.23 | 19.56 ± 6.38 | U/L |
| | Aspartate aminotransferase (AST) | 26.46 ± 10.90 | 27.69 ± 11.82 | 23.92 ± 4.67 | U/L |
| Renal Parameters | Cystatin C | 0.94 ± 0.21 | 0.97 ± 0.24 | 0.83 ± 0.08 | mg/L |
| | Creatinine | 73.55 ± 22.45 | 75.75 ± 25.87 | 68.72 ± 10.09 | µmol/L |
| | Total protein | 72.22 ± 4.24 | 72.86 ± 4.24 | 71.96 ± 3.33 | g/L |
| | Urea | 5.57 ± 1.52 | 5.72 ± 1.67 | 5.13 ± 1.09 | mmol/L |
| | Phosphate | 1.16 ± 0.16 | 1.15 ± 0.16 | 1.15 ± 0.14 | mmol/L |
| | Urate | 313.88 ± 81.12 | 336.99 ± 83.75 | 280.45 ± 58.31 | µmol/L |
| Diabetes Parameters | Hemoglobin A1c (HbA1c) | 36.60 ± 6.74 | 38.14 ± 8.41 | 34.07 ± 3.16 | mmol/mol |
| | Glucose | 5.17 ± 1.26 | 5.38 ± 1.59 | 4.87 ± 0.57 | mmol/L |
| Bone and Joint | Vitamin D | 50.49 ± 21.37 | 47.68 ± 21.09 | 50.69 ± 20.89 | nmol/L |
| | Rheumatoid factor | 24.87 ± 19.62 | 24.91 ± 20.28 | 24.42 ± 20.92 | IU/ml |
| | Alkaline phosphatase (ALP) | 85.73 ± 29.51 | 87.09 ± 26.96 | 76.48 ± 17.71 | U/L |
| | Calcium | 2.38 ± 0.10 | 2.39 ± 0.10 | 2.37 ± 0.07 | mmol/L |
| Cancer Parameters | Sex Hormone-Binding Globulin (SHBG) | 52.48 ± 27.42 | 46.03 ± 24.21 | 53.69 ± 23.28 | nmol/L |
| | Testosterone | 6.63 ± 6.02 | 7.06 ± 5.76 | 6.64 ± 6.38 | nmol/L |
| | Oestradiol | 421.50 ± 422.81 | 375.81 ± 339.37 | 478.64 ± 377.52 | pmol/L |
| | Insulin-like Growth Factor 1 (IGF 1) | 20.96 ± 5.80 | 20.71 ± 5.87 | 22.09 ± 4.50 | nmol/L |
| Complete Blood Count (CBC) | White Blood Cell count (WBC) | 6.96 ± 2.09 | 6.97 ± 2.28 | 6.80 ± 1.27 | x10^9/L |
| | Red Blood Cell count (RBC) | 4.51 ± 0.42 | 4.50 ± 0.42 | 4.58 ± 0.31 | x10^9/L |
| | Haemoglobin (Hgb) | 14.14 ± 1.24 | 14.14 ± 1.25 | 14.15 ± 0.92 | g/dL |
| | Hematocrit (Hct) | 40.99 ± 3.54 | 40.99 ± 3.56 | 41.14 ± 2.65 | % |
| | Mean Corpuscular Volume (MCV) | 91.15 ± 4.64 | 91.16 ± 4.61 | 89.97 ± 2.93 | fL |
| | Mean Corpuscular Hemoglobin (MCH) | 31.46 ± 1.95 | 31.47 ± 1.95 | 30.94 ± 1.08 | pg |
| | Mean Corpuscular Hemoglobin Concentration (MCHC) | 34.51 ± 1.12 | 34.52 ± 1.12 | 34.39 ± 0.54 | g/dL |

| | | | | |
|---|---|---|---|---|
| Platelet count (Plt) | 253.33 ± 61.94 | 253.06 ± 61.55 | 255.03 ± 47.03 | x10^9/L |
| Mean Platelet Volume (MPV) | 0.23 ± 0.05 | 0.23 ± 0.05 | 0.23 ± 0.04 | fL |
| Lymphocyte count (Absolute) | 1.97 ± 1.17 | 1.97 ± 1.15 | 2.01 ± 0.51 | x10^9/L |
| Monocyte count (Absolute) | 0.48 ± 0.22 | 0.48 ± 0.40 | 0.49 ± 0.13 | x10^9/L |
| Neutrophil count (Absolute) | 4.29 ± 1.44 | 4.30 ± 1.46 | 4.08 ± 0.94 | x10^9/L |
| Eosinophil count (Absolute) | 0.18 ± 0.14 | 0.18 ± 0.14 | 0.18 ± 0.10 | x10^9/L |
| Basophil count (Absolute) | 0.03 ± 0.05 | 0.03 ± 0.05 | 0.04 ± 0.02 | x10^9/L |
| Lymphocyte percentage | 28.70 ± 7.59 | 28.67 ± 7.55 | 29.66 ± 5.76 | % |
| Monocyte percentage | 7.03 ± 2.79 | 7.04 ± 2.80 | 7.33 ± 1.62 | % |
| Neutrophil percentage | 61.11 ± 8.67 | 61.13 ± 8.60 | 59.82 ± 6.19 | % |
| Eosinophil percentage | 2.59 ± 1.88 | 2.59 ± 1.86 | 2.63 ± 1.43 | % |
| Basophil percentage | 0.57 ± 0.63 | 0.57 ± 0.63 | 0.57 ± 0.27 | % |
| Neutrophil to Lymphocyte Ratio (NLR) | 2.40 ± 1.30 | 2.40 ± 1.27 | 2.13 ± 0.63 | N/A |
| Lymphocyte to Monocyte Ratio (LMR) | 4.65 ± 3.36 | 4.64 ± 3.78 | 4.24 ± 1.23 | N/A |
| Platelet to Lymphocyte Ratio (PLR) | 141.99 ± 60.39 | 141.88 ± 61.42 | 134.59 ± 40.11 | N/A |

**Table S6.** Mean and standard deviation of demographics and blood test parameters in cancer group, cardiovascular group and final control group subjects

| Cancer Type | Learning Rate | Maximum Depth | Number of estimators |
|---|---|---|---|
| Prevalent Solid Cancer | 0.1 | 3 | 100 |
| Prevalent Blood Cancer | 0.1 | 3 | 1000 |
| Prevalent Benign Cancer | 0.01 | 7 | 1000 |
| Incident Solid Cancer | 0.05 | 9 | 1000 |
| Incident Blood Cancer | 0.01 | 7 | 1000 |
| Incident Benign Cancer | 0.01 | 3 | 1000 |

**Table S7.** Per label tuned hyperparameters for the final model

The tuned hyperparameters for the final model were provided above. In general, a greater number of estimators (1000 estimators) was required for cancer prediction of most cancer types.

**Feature Importance Analysis**

| Disease / Cancer Type | Prevalent — Top Predictors (SHAP Importance) | Incident — Top Predictors (SHAP Importance) |
|---|---|---|
| **Solid Cancer** | Age ↑, Sex (Male ↑), Cystatin C ↑, Basophil Count ↓, LDL direct ↓, HDL cholesterol ↓ | Age ↑, Sex (Male ↑), Cystatin C ↑, Basophil Count ↓, Triglycerides ↑, C-reactive protein ↑ |
| **Blood Cancer** | Cystatin C ↑, Age ↑, Apolipoprotein B ↑, CRP ↑, HDL cholesterol ↓ | Cystatin C ↑, Age ↑, Sex (Male ↑), Monocyte Count ↑, Triglycerides ↑ |
| **Benign Cancer** | Age ↑, Basophil Count ↓, Cystatin C ↑, LDL direct ↓, Apolipoprotein A ↓ | Age ↑, Basophil Count ↓, Triglycerides ↑, Cystatin C ↑, CRP ↑ |
| **Colorectal Cancer** | Age ↑, Sex (Male ↑), Cystatin C ↑, CRP ↑, Hemoglobin A1c ↑, Neutrophil-to-Lymphocyte Ratio ↑ | Age ↑, Sex (Male ↑), Cystatin C ↑, Glucose ↑, Triglycerides ↑, Basophil Count ↓ |
| **Cardiovascular Disease** | Age ↑, Sex (Male ↑), Cystatin C ↑, LDL direct ↑, HDL cholesterol ↓, Triglycerides ↑ | Age ↑, Sex (Male ↑), Systolic BP ↑, Cystatin C ↑, CRP ↑, Apolipoprotein B ↑ |

Table S8. SHAP Summary of feature Improtance Key Predictors for Cancer and Cardiovascular Disease. ↑ = positive association with disease risk | ↓ = inverse association.

Across all disease models, age, sex, and Cystatin C consistently dominated as global predictors, reflecting demographic and metabolic influences on disease susceptibility.

- Basophil count inversely correlated with multiple cancer categories, highlighting immune-hematologic suppression in malignancy.
- Lipid-related markers (LDL, HDL, Triglycerides, Apolipoproteins A/B) contributed prominently to both cancer and cardiovascular predictions, emphasizing shared metabolic-inflammatory axes.
- CRP and Hemoglobin A1c emerged as inflammatory-metabolic signatures particularly strong for colorectal and cardiovascular risk stratification.

- SHAP visualizations for these relationships and per-feature value distributions are provided in the *Supplementary Information*.

Collectively, these findings indicate that integrating hematologic, renal, and lipidomic biomarkers, especially Cystatin C, CRP, and lipid ratios, within digital-twin frameworks can enhance both cancer and cardiovascular disease screening precision.

| **Cancer Type** | **Learning Rate** | **Maximum Depth** | **Number of estimators** |
|---|---|---|---|
| Has Colorectal Cancer | 0.1 | 5 | 1000 |
| Prevalent Colorectal Cancer | 0.01 | 5 | 500 |
| Incident Colorectal Cancer | 0.01 | 5 | 500 |

**Table S9.** Per label tuned hyperparameters for colorectal cancer prediction model

| **Cardiovascular disease** | **Learning Rate** | **Maximum Depth** | **Number of estimators** |
|---|---|---|---|
| Angina | 0.01 | 5 | 1000 |
| Heart Attack | 0.01 | 5 | 1000 |
| Hypertension (high bp) | 0.01 | 7 | 1000 |
| Stroke | 0.01 | 5 | 500 |

**Table S10.** Per-label tuned hyperparameters for cardiovascular disease prediction

The optimal learning rate for all cardiovascular disease prediction models was 0.01. The optimal maximum depth for most cardiovascular diseases (except hypertension) was 5 and the optimal number of estimators for most cardiovascular diseases (except stroke) was 1,000.